\documentclass[useAMS,usenatbib]{mn2e}

\usepackage{graphicx,epsfig,subfigure}
\usepackage{url}

\usepackage{amsmath}
\usepackage{amssymb}

\usepackage{threeparttable}
\usepackage{float}

\title[Dark matter in early-type galaxies]
   {How much dark matter is there inside early-type galaxies?\thanks{Based on data taken
    from the DR9 of the Sloan Digital Sky Survey.}}

    \author[Nigoche-Netro et al.]
  {A. Nigoche-Netro,$^1$\thanks{E-mail: anigoche@gmail.com}
  A. Ruelas-Mayorga,$^2$ P. Lagos,$^3$ G. Ramos-Larios,$^1$
  \newauthor 
    C. Kehrig,$^4$
    S. N. Kemp,$^1$
    A. D. Montero-Dorta,$^4$
    and J. Gonz\'alez-Cervantes$^1$ \\
 $^1$Instituto de Astronom{\'\i}a y Meteorolog{\'\i}a, Universidad
     de Guadalajara, Guadalajara, Jal. 44130, M\'exico\\
 $^2$Instituto de Astronom\'{\i}a, Universidad Nacional Aut\'onoma
     de M\'exico, Cd. Universitaria, M\'exico, D.F. 04510, M\'exico\\
 $^3$Centro de Astrofisica da Universidade do Porto, Rua das Estrelas, 4150-762 Porto, Portugal.\\
 $^4$Instituto de Astrof\'isica de Andaluc\'ia (IAA), Glorieta de la Astronom\'ia s/n, 18008, Granada, Spain.}

\begin{document}

\date{Accepted 2014. Received 2013 February; in original form 2013 February}

\pagerange{\pageref{firstpage}--\pageref{lastpage}}
\pubyear{2014}

\maketitle

\label{firstpage}


\begin{abstract}
 We study the luminous mass as a function of the
dynamical mass inside the effective radius ($r_{e}$) of early-type galaxies (ETGs) to search for differences
between these masses. We assume Newtonian dynamics and that any difference between these masses is due
to the presence of dark matter. We use several samples of ETGs -ranging from 19 000 to 98 000 objects- from the ninth data release of the Sloan
Digital Sky Survey. We perform Monte Carlo (MC) simulations of galaxy samples and compare them with real samples.
The main results are:  i) MC simulations show that the distribution of the dynamical vs. luminous mass
depends on the mass range where the ETGs are distributed (geometric effect). This dependence is
caused by selection effects and intrinsic properties of the ETGs. ii) The amount of dark matter inside $r_{e}$ is
approximately $7\%$ $\pm$ 22\%. iii) This amount of dark matter is lower than the minimum estimate (10\%)
found in the literature and four times lower than the average (30\%) of literature estimates. However, if we consider the associated error, our estimate is of the order of the literature average.

\end{abstract}




\section{Introduction}
\label{sec:intro}

The masses of galaxies have traditionally been measured either by estimating their stellar, gas and dust content \citep{spi75}, or, for spiral galaxies, by measuring their rotation velocities at different radii, or by measuring their velocity dispersion in the case of elliptical or irregular galaxies \citep{bur75,sof01,sim07}.

The estimation of the stellar, gas and dust content of galaxies is usually carried out by using the amount of radiation we detect coming from them, and invoking typical mass to light ratios (${\bf M}$/$L$) that have been calibrated using different stellar samples in our own Galaxy. Measuring the rotation velocity of spiral galaxies at different radii produces the rotation velocity curve. These rotation velocity curves are tools which are used, among other things, for determining the amount and distribution of mass interior to a given radius, to study the kinematics of galaxies, to derive an insight into their evolutionary histories and the possible role that interactions with other systems have played. Departures from the expected rotation curves may be related to the possible presence and distribution of dark matter. Rotation curves are also very useful since they may be obtained at different wavelengths and, therefore, provide information as to the kinematics of different constituents of the galaxy. Rotation curves may be observed in the optical and infrared to trace the stellar and ionized gas motions, as well as in the radio and microwave regimes which  trace the neutral and molecular gas components of a galaxy. It was precisely due to the fact that many rotation curves of spirals turned out to be flat towards the galactic outskirts that the presence of a yet unknown type of mass, now denoted as dark matter, was suggested. See \cite{sof01}, and references therein.

More recently, galactic masses have also been measured by means of stellar population synthesis models which give an idea of the total stellar content of a galaxy as well as the distribution of stars of all the different spectral types and luminosity classes \citep{aug10,bar11,son12}. These methods assume a universal IMF, however, recently it has been demonstrated that the IMF is not universal and that it depends on ${\bf M}$/$L$ \citep{cap12}.

Galactic masses have also been calculated using dynamical models of ETGs, for example \cite{vandermarel1991} constructed dynamical models for 37 bright elliptical galaxies. From these models he found an average (M/$L)_B=(5.93 \pm 0.25)h_{50}$. Discrepancies of the observed velocities in the outer parts of the galaxies with those predicted by the models implied that no axisymmetric models with constant mass-to-light ratios appropriately described the kinematical behaviour of these galaxies in their outskirts, these results may indicate the presence of dark haloes. Dynamical models made by \cite{magorrian1998} for 36 nearby galaxies observed with the HST and ground-based telescopes indicated that in 30 of these galaxies, it was necessary to invoke the presence of a central massive dark object (MDO) whose mass amounted to M$_{\bullet} \sim 0.006$M$_{bulge}$. They concluded that $\sim 97 \%$ of ETGs had MDOs whose masses distributed normally in log(M$_{\bullet}/$M$_{bulge})$ with a mean value of $-2.28$.

A detailed study of the pure disk galaxy $M33$ led \cite{gebhardt2001} to conclude that supermassive black holes (SMBH) were associated with galaxy bulges and not with their disks and that the SMBH mass correlated with the velocity dispersion of the bulges. Dynamical studies of line-profile shapes of 21 elliptical galaxies were used by \cite{gerhard2001} to investigate the dynamical family relations and dark halo properties of these galaxies. They found interesting results regarding their circular velocity, radial anisotropy, the fact that ETGs also obeyed a Tully-Fisher relation, etc. They also found that elliptical galaxies have nearly maximal M/$L_B$ which implied minimal haloes. As far as the presence of dark matter was concerned, some galaxies showed no dark matter within $2r_e$, whereas others presented values of M/$L_B$ of $20-30$ at $2r_e$. Their maximum stellar mass models required $\sim 10\%-40\%$ of dark matter within $r_e$, and equal amounts of dark and luminous matter were needed at $\sim 2-4r_e$. Using two-integral Jeans and three-integral Schwarzschild dynamical models of 25 elliptical and lenticular galaxies, \cite{cap06} studied the correlations between the dynamical mass-to-light ratios M/$L$ and other global observables. They found no difference between the M/$L$ ratio of galaxies in the field and in clusters. They determined that the amount of dark matter inside $r_e$ was $\sim 30\%$ of the total mass contained there. They speculated that the amount of dark matter appeared to correlate with the galactic rotation velocity; the fast-rotating galaxies contain less dark matter compared with the more massive slow-rotating ones.

\cite{thomas2007} made dynamical models of 17 early-type galaxies in the Coma cluster. Their best model fit including dark matter matches the data better than the best fit without dark matter. They found that the central phase-space density of dark matter is lower $(1/10)$ than that of luminous matter. \cite{williams2009} produced mass models for a sample of 14 spiral and 14 S0 galaxies in order to constrain their stellar and dark matter content. These mass models contain within the effective radius $r_e$ a median $15\%$ dark matter. Dark and stellar matter contribute equally in a $4.1r_e$ radius sphere. \cite{thomas2011} made dynamical models of galaxies in the Coma cluster. Assuming that the mass distribution is equal to the light distribution, they obtained dynamical masses that do not agree with those obtained from strong gravitational lens systems of similar velocity dispersion. To make them match it is necessary to include, in the dynamical fits, dark matter haloes. A constant IMF proves inadequate because the amount of mass that follows the light increases more rapidly with the velocity dispersion than expected. They conclude that a variation of the IMF with the galactic velocity dispersion may correct this problem. For galaxies containing old stellar populations it appears that a systematic variation of the total (stellar + dark) mass within the effective radius explains the tilt of the fundamental plane. \cite{cap13} made axisymmetric dynamical models for $260$ galaxies in the $ATLAS^{3D}$ sample. They derived accurate mass-to-light ratios (M/$L)_e$ and dark matter fractions $f_{DM}$, in a sphere of radius $r_e$. Their $f_{DM}$ distribution presented a median value of $13\%$ in their galaxy sample.

Masses of galaxies have also been measured  using the gravitational lens phenomenon which provides precise and direct measurements of galactic masses at different scales. Measurements using the gravitational lens phenomenon also allow us to establish the presence and nature of dark matter. For a number of years it has been believed that elliptical galaxies are surrounded by extended haloes of dark matter \citep{tre10} that follow the \cite{nav96} density profiles, although it appears in some cases that little dark matter is present in the inner $(r \leq r_e)$ parts of the galaxy, whereas in other more recent studies it seems that up to $30\%$ of the mass content inside one effective radius is dark matter \citep{tre04}. Studying the kinematics of stars, globular clusters, and cold and hot gas in nearby systems \citep{ber93,hum06}, it has been established that dark matter haloes are required to explain the dynamics of massive elliptical galaxies, as long as we require Newtonian gravity to be valid at these scales.

Weak-lensing observations have been used to reveal the presence of and to characterise dark matter haloes of elliptical galaxies out to redshifts $z \sim 0.5$ \citep{lag10,hoe05,gav07}. For ETGs with redshifts in the interval $0.5 \leq z \leq 1$, the presence of dark matter haloes has been suggested by strong-lensing observations.

The relative spatial distribution of luminous and dark matter is studied through the determination of the fraction of total mass in the form of dark matter within a projected fraction or multiple of the effective radius \citep{jia07}. $f_{DM}$ appears to increase with growing radius reaching values of $\sim 70\%$ at five effective radii \citep{tre04}. Furthermore, $f_{DM}$ within a fixed radius seems to grow with galaxy stellar mass and with velocity dispersion \citep{tor09,nap10,gra10,aug10a}.


$f_{DM}$ ultimately reveals the degree of difference between dynamical mass and luminous mass in a galaxy. $f_{DM}$ may vary from very large values, as has been found for dwarf spheroidal galaxies by \cite{sim07}, to small values as in the case of bright giant elliptical galaxies \citep{rom03}. Studies of the Virgo giant elliptical galaxy NGC 4949 (M60) by \cite{teo11} reveal that the kinematics of Planetary Nebulae in this object is consistent with the presence of a dark matter halo with $f_{DM} \sim 0.5$ for $r=3r_e$. \cite{deb01} presented three-integral axisymmetric models for NGC 4649 and NGC 7097 and concluded that the kinematic data for NGC 4649 only require a small amount of dark matter, however \cite{das11} determine $f_{DM} \sim 0.78$ at $r=4r_e$ for NGC 4649. \cite{koopmans2006} present a gravitational lensing and stellar-dynamical analysis of 15 massive field ETGs. They find an average projected dark matter fraction of $<f_{DM}>=0.25 \pm 0.06$ inside the average Einstein radius assuming the values of stellar mass-to-light ratios from the fundamental plane. The average inner mass density of the galaxies studied presents no evolutionary changes suggesting a collisional scenario that leads to a dynamically isothermal mass distribution. \cite{bar11} study the internal mass distribution, the amount of dark matter and the dynamical structure of sixteen early-type lens galaxies from the SLACS survey at $z=0.08-0.33$. By combining the constraints from gravitational lensing and stellar kinematics, they determine, among other things, the lower limit for dark matter $f_{DM}$ inside the effective radius. This fraction varies from almost zero to almost $50 \%$ with a median value of $12 \%$. Including stellar population synthesis models and using a Salpeter IMF they found an average $f_{DM}$ inside one effective radius of $\sim 0.31$. This value increases to $f_{DM} \sim 0.61$ if a Chabrier IMF is assumed instead. Finally \cite{nig11} comparing masses obtained from colours and masses obtained via the virial theorem for approximately 90 000 ETGs found an almost negligible amount of dark matter inside $r_e$.

It is well known that direct detection of dark matter has not been achieved yet. Its presence has been deduced from dynamical analysis in which Newtonian gravity is required to be valid. Alternatively, it may be thought that the current Newtonian and general relativistic theories of gravity fail at these very low acceleration regimes. A modification of Newtonian dynamics (MOND) \citep{mil83} has been proposed and further developments along these lines have been able to explain a variety of phenomena without the need to invoke the presence of dark matter e.g. spiral galaxies flat-rotation curves \citep{san02}, projected surface density profiles and observational parameters of the local dwarf spheroidal galaxies \citep{her10,mcg10,kro10}, the relative velocity of wide binaries in the solar neighbourhood \citep{her12}, fully self-consistent equilibrium models for NGC 4649 \citep{jim13} and references in these papers among others.

In this work we present a study of luminous and dynamical mass inside the effective radius of ETGs considering Newtonian dynamics. We search for differences between these masses and assume that any difference is due to dark matter. It is important to stress at this point that whenever we quote values for the amount of dark matter, these correspond to mean values calculated in different intervals of mass depending on the case being discussed at the time.

The structure of this study is as follows; in \S 2 we present the sample of ETGs we use in this work, in \S 3 we calculate the virial and stellar masses for all the galaxies in the sample, \S 4 discusses the amount of stellar mass as a function of virial mass, \S 5 deals with the amount of dark matter found inside ETGs, and finally, \S 6 presents our conclusions.


\section{The sample of ETGs}

We use a sample of ETGs from the Ninth Data Release (DR9) of the Sloan Digital Sky Survey (SDSS) (\citealt{yor00}; \citealt{aba09}; \citealt{aih11}) in the $g$ and $r$ filters. This sample contains approximately 98000 galaxies in each filter, distributed in a redshift interval $0.0024 < z < 0.3500$ and within a magnitude range $<\Delta M>$ $\sim 7$ $mag$ ($-17.5 \ge M_{g} > -24.5$). This sample shall be called hereafter, ``Total-SDSS-Sample". The selection criteria of the Total-SDSS-Sample are similar to those used in \citet{hyd09} and \citet{nig10}; i.e.:


1) The brightness profile of the galaxy must be well adjusted by a de Vaucouleurs profile, in both the $g$ and $r$ filters (fracdevg = 1 and fracdevr = 1 according to the SDSS nomenclature).

2) The de Vaucouleurs magnitude of the galaxies must be contained in the interval $14.5 < m_{r, dev} < 17.5$ and its equivalent in the g filter.

3) The quotient of the semi axes (b/a) for the galaxies must be larger than 0.6 in both filters $g$ and $r$.

4) The galaxies must have a velocity dispersion of $\sigma_0 >$ 60 $km/s$ and a signal-to-noise ratio (S/N) $>$ 10.


The fundamental difference between the selection procedure in \citet{nig10}, and that of the present paper, is that in this work we do not use the morphological parameter {\it eclass}, since it is not given in the DR9. {\it eclass} means early-type spectrum in the SDSS nomenclature. An analysis performed on a sample of approximately 90 000 ETGs from the DR7 in \citet{nig10} indicates that there is a difference of 21 galaxies when applying the {\it eclass} criterion from the number obtained when this parameter is not used. This demonstrates that the rest of the criteria used are capable of obtaining the same selection of ETGs, and that the exclusion of the {\it eclass} parameter will not produce a significant change in the number of galaxies found in the sample drawn from the DR9. In the DR9 we find a new morphological classification, which appears to be more rigorous than the {\it eclass} parameter. This new morphological classification is obtained from the Galaxy Zoo project (see \citealt{lin08}) and can be found in the Zoospec catalogue in the DR9. In this catalogue we may only find those galaxies for which there is spectroscopic information from the DR7. The total number of ETGs considering only the morphological classification from Zoospec is approximately 59,000. If in addition, we require this sample to fulfill the criteria 1-4 listed above, the sample is reduced to approximately 27,000 ETGs. This last sample shall be referred to as ``The-Morphological-Sample". Given that the aim of this work is making use of the new photometric and spectroscopic information in the DR9, The-Morphological-Sample will only  be used to perform comparisons with the Total-SDSS-Sample.

In addition, using the general restrictions 1-4 defined above, we extract a volume-limited sample (0.04 $\leq\;z\;\leq$ 0.08) of approximately 19 000 ETGs from the DR9 in the g and r-band filters. This subsample covers a magnitude range $<\Delta M>$ $\sim 4.5$ $mag$ ($-18.5 \ge M_{g} > -23.0$) in both filters and is approximately complete for $M_{g} \leq -20.0$ (see \citealt{nig10}; \citealt{nig11} for details). We shall refer to it as ``The-Homogeneous-SDSS-Sample".

The photometry and spectroscopy of the samples of galaxies drawn from the DR9 require a series of corrections that are listed as follows:

\begin{itemize}

\item Seeing correction: We use the seeing-corrected parameters (total magnitude and effective radius) from the SDSS pipeline.

\item Extinction correction: We use the extinction correction values from the SDSS pipeline.

\item K correction: We use the K correction values from Bernardi et al. (2003a) and apply them to our sample as follows:

\begin{equation}
k_{g}(z)\, \; =\, \; -5.261\; z^{1.197},
\end{equation}

\begin{equation}
k_{r}(z)\, \; =\, \; -1.271\; z^{1.023}.
\end{equation}

\item Cosmological dimming correction: We use the cosmological dimming correction of J\o rgensen et al. (1995a).

\item Evolution correction: Bernardi et al. (2003b) report that the more distant galaxies in their sample are brighter than those nearby.
We use their results and apply them to our sample of galaxies as follows:

\begin{equation}
ev_{g}(z)\, \; =\, \; +1.15\; z,
\end{equation}

\begin{equation}
ev_{r}(z)\, \; =\, \; +0.85\; z.
\end{equation}

\item Effective radius correction to the rest reference frame: Given that ETGs have colour gradients, their mean effective radii at longer wavelengths are smaller. To correct for this effect, we follow the procedure given in Hyde \& Bernardi (2009).

\item Aperture correction to the velocity dispersion: The velocity dispersion for ETGs appears to have radial gradients. This means that the velocity dispersion values given in the SDSS ($\sigma_{SDSS}$) depend on both the distance of the object and the size of the aperture used for the observations ($r_{ap}$). To correct our data to a system that is independent of both the distance and instrument used for the observations, we use an expression derived by J\o rgensen et al. (1995b) as follows:

\begin{equation}
log\left(\frac{\sigma_{SDSS}}{\sigma_{e}}\right)\, \; =\, \; -0.065 {log\left(\frac{r_{ap}}{r_{e}}\right)}-0.013 \left[log\left(\frac{r_{ap}}{r_{e}}\right)\right]^{2},
\end{equation}

where $r_{e}$ is the effective radius in arcsec and $\sigma_{e}$ is the corrected velocity dispersion, that is to say, the velocity dispersion inside $r_{e}$. For the SDSS case $r_{ap}$=$1.5$ $arcsec$.

\end{itemize}

 \section{Calculation of the stellar and virial mass of the ETGs}

In this paper we use two methods to derive the mass of galaxies. The first one makes use of the luminosity of the galaxies, the mass calculated in this way will be named ``the stellar mass". The second method makes use of Newtonian dynamics, the mass obtained in this way will be named ``the virial mass". In the following sections we shall discuss in detail these methods.

\subsection{The stellar mass}

\subsubsection{The stellar mass considering an universal IMF}

In what follows we will describe three different estimations of the stellar mass considering an universal IMF.

In the first two cases we make use of an equation for stellar mass-to-light (M/L) ratios obtained from fits of optical and near-infrared galaxy data with simple stellar population synthesis models and considering an universal Salpeter IMF (\citealt{bel03}). The equation is as follows:

\[
{\bf
M_{g} \sim L_{g}} 10^{a_{g} + b_{g} (\it M_{g}-M_{r})}, \;\;\;\;\;\;\;\;\;\;\;\;\;\;\;\;\;\;\;\;\;\;\;\;\;\;\;\;\;\;\;\;\;\;\;\;\;\;\;\;\;\;\;\;\;6
\]

where ${\bf M_{g}}$ is the total stellar mass in the $g$ filter, ${\bf L_{g}}$ is the luminosity in the $g$ filter, $M_{g}$ and $M_{r}$ are the magnitudes in the $g$ and $r$ filters, and $a_{g}$ and $b_{g}$ are scale factors (see Table 7 from \citealt{bel03}).

i) de Vaucouleurs Salpeter-IMF stellar mass. The masses have been derived using equation 6 and model-parametric photometric information from the SDSS-DR9. That is to say, the magnitude and effective radius of the ETGs have been obtained using a de Vaucouleurs galaxy light profile.

ii) S\'ersic Salpeter-IMF stellar mass. The masses have been obtained considering equation 6 and S\'ersic parameters. The S\'ersic parameters have been obtained from the SDSS-DR9 Petrosian parameters following \cite{gra05}. These mass values, which shall be referred to  as S\'ersic masses, will be used to find a possible bias on our results due to the luminous profile fit (See sections 3.4 and 5).

iii) Kroupa-IMF stellar mass.  In this case the stellar masses were derived by the MPA-JHU group ($http://www.mpa-garching.mpg.de/SDSS/$)\citep{kau03,bri04,trm04}. They estimate the stellar mass within the SDSS spectroscopic fibre aperture using fibre magnitudes and the total stellar mass using model magnitudes. These stellar masses are calculated using the Bayesian methodology and model grids described in \cite{kau03}. A Kroupa (2001) initial mass function is assumed. These stellar masses can be obtained directly from the MPA-JHU group or from the GALSPEC catalogue of the SDSS-DR9 ($http://www.sdss3.org/dr9/algorithms/galaxy_mpa_jhu.php$).

According to \cite{sch10}, the stellar mass inside a sphere of radius $r_{e}$ corresponds approximately to 42\% of the total stellar mass calculated using the procedures mentioned above.

\subsubsection{Correction to the stellar mass due to a non-universal IMF}

 The stellar masses described in the previous section assume a universal (Salpeter or Kroupa) IMF, however recent works have found that the IMF is not universal, for example, \cite{cap12} and \cite{dut13} show that the IMF depends on the stellar mass. Considering that the mass follows light, using $\Lambda$CDM models that reproduce the relations between galaxy size, light concentrations and stellar mass and using the spherical Jeans equations to predict velocity dispersion of a sample of ETGs, \cite{dut13} propose a correction to the universal stellar mass as follows:

 \[
 log \left [\frac{{\bf M_{star}}}{{\bf M_{uni}}}\right] \sim a +b\;log \left [\frac{{\bf M_{uni}}}{10^{11} {\bf M_{\odot}}}\right], \;\;\;\;\;\;\;\;\;\;\;\;\;\;\;\;\;\;\;\;\;\;\;\;\;\;\;\;\; 7
\]				                                           		

where ${\bf M_{star}}$ and ${\bf M_{uni}}$ are the stellar corrected-IMF mass and the stellar universal-IMF mass respectively and $a$, $b$ are scale factors.

The sample of ETGs used in \cite{dut13} has several similarities to our total sample, for example, the mass range and the redshift range are approximately equal to those used by us. The main differences are the source of the sample and the total number of ETGs that it contains. While we use the SDSS-9DR they use the SDSS-7DR. In the case of the total number of galaxies, it differs because, although the selection criteria are similar, some of them are more lax. For example, we use a limit in the quotient of the semi axes of b/a $>$ 0.6 while they use b/a $>$ 0.5, and this different limit makes a big difference in the number of galaxies included.

We shall use the correction proposed by Dutton (Equation 7) to the stellar mass of our samples. It is important to say that equation 7 was obtained considering a Chabrier IMF \citep{cha03} so in order to apply it correctly we have converted our Salpeter and Kroupa IMF masses to the Chabrier IMF mass by subtracting 0.23 dex and 0.035 dex respectively \citep{dut13}. In section 5 we shall present and discuss the results.

\subsection{The virial mass}

This method requires the value of the velocity dispersion, and assumes that the galaxies are in virial equilibrium. It uses the following equation:

\[
{\bf M_{virial}} \sim  K \frac{ r_{e} \sigma_{e}^{2}}{G}, \;\;\;\;\;\;\;\;\;\;\;\;\;\;\;\;\;\;\;\;\;\;\;\;\;\;\;\;\;\;\;\;\;\;\;\;\;\;\;\;\;\;\;\;\;\; \;\;\;\;\;\;\;\;\;\;\;\;\;8
\]

 where ${\bf M_{virial}}$ is the virial mass, $r_{e}$ is the effective radius, $\sigma_{e}$ is the velocity dispersion inside $r_{e}$, $G$ is the gravitational constant and $K$ is a scale factor that for the case of the de Vaucouleurs profile takes the value of 5.953 \citep{cap06}.

The mass calculated using equation 8 gives the approximate value of the dynamical mass of the galaxies inside a sphere of radius equal to the effective radius $r_{e}$ (see \citealt{cap06,sch10}). This mass may be luminous or not.

In sections 4 and 5 we will perform an analysis of the behaviour of the masses calculated by means of both methods listed above. We shall only consider the region internal to $r_{e}$.


\begin{figure*}

  \begin{center}



  \includegraphics[width=18cm]{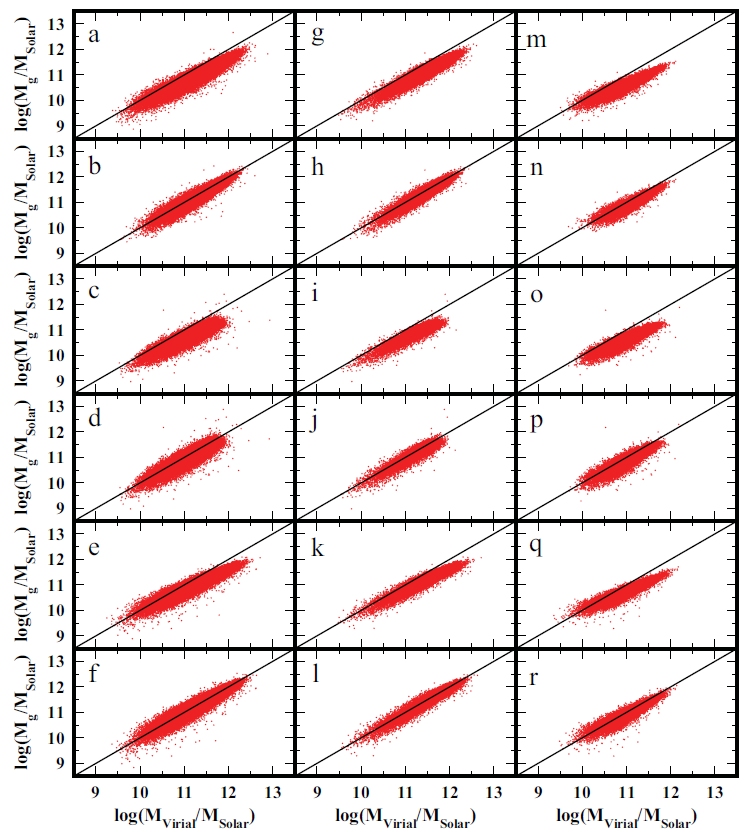}

      \caption{Distribution of the stellar mass with respect to the virial mass for the ETGs Total (column 1), Morphologic (column 2) and Homogeneous (column 3) samples. The first row corresponds to the Salpeter IMF stellar mass, the second one corresponds to the Salpeter IMF-corrected stellar mass, the third one corresponds to the S\'ersic Salpeter stellar mass, the fourth one corresponds to the S\'ersic Salpeter IMF-corrected stellar mass, the fifth one corresponds to the Kroupa IMF stellar mass and the last one corresponds to the Kroupa IMF-corrected stellar mass. The solid line is the one-to-one line.}
         \label{FigVibStab}

  \end{center}

\end{figure*}

\subsection{Errors in the virial and stellar masses}

A key factor to calculate the error in the masses of the ETGs is the error associated with the parameters included in its calculation. To estimate the de Vaucouleurs Salpeter-IMF stellar mass we use the model-parametric photometric information, as well as its associated errors, from the SDSS-D9R. On the other hand, the S\'ersic Salpeter-IMF mass and the Kroupa-IMF stellar mass have been obtained considering the Petrosian magnitude and Petrosian radius from the SDSS-DR9. The Petrosian radius is defined as the largest radius at which the $r$-band surface brightness is at least one-fifth the mean surface brightness interior to that radius. The Petrosian flux is the total flux within a circular aperture of twice the Petrosian radius. In the case of the virial mass, besides the effective radius we require the velocity dispersion, which has been calculated inside the radius subtended by the SDSS fibre. These parameters have been corrected for various sources of bias (see Section 2) and the final errors have been obtained considering the rules of error propagation. The mean errors of the photometric and spectroscopic parameters from the SDSS data give the following results:

\begin{itemize}

\item The mean error in the de Vaucouleurs magnitude ($g$ filter) is approximately 0.008 mag.

\item The mean error in the Petrosian magnitude ($g$ filter) is approximately 0.02 mag.

\item The mean error in the de Vaucouleurs effective radius ($g$ filter) is of the order 4\%.

\item The mean error in the Petrosian radius ($g$ filter) is of the order 6\%.

\item The mean error in the velocity dispersion is approximately 5\%.

\end{itemize}

However, there are some recent studies that show that the errors in these parameters could be underestimates. For example, using detailed axisymmetric dynamical models, \cite{cap13} show that the photometric parameters depend, among other things, on the adopted galaxy light profile, the extrapolation of the outermost part of the galaxy light profile and the depth of the photometry used. Taking into account these systematics they find that the magnitude ($r$ filter) as well as the effective radius are accurate at the 10\% level. On the other hand, using a model-independent approach \cite{che10} find that the B magnitude mean error is approximately 0.13 mag and the effective radius error is approximately 10\%. In the case of the spectroscopic parameter $\sigma_{e}$, its accuracy is affected mainly by uncertainties in the stellar templates. The general finding is that this parameter can be reproduced at best with an accuracy of 5\% \citep{cap12,ems04}.

To take into account possible systematic errors in the structural parameters of our ETGs sample, we adopt more conservative mean errors considering the previously mentioned studies \citep{cap13,che10,ems04}, of 0.1 mag, 10\% and 5\% for the magnitude, effective radius and velocity dispersion respectively. Considering these data, the mean errors for the stellar and virial mass are approximately 15\% and 12\% respectively.

 \section{Distribution of the stellar mass as a function of the virial mass of ETGs}

In Figure 1 we show a mosaic of the distribution of the stellar mass with respect to the virial mass for the ETGs Total, Morphologic and Homogeneous samples. The first row corresponds to the Salpeter IMF stellar mass, the second one corresponds to the Salpeter IMF-corrected stellar mass, the third one corresponds to the S\'ersic Salpeter stellar mass, the fourth one corresponds to the S\'ersic Salpeter IMF-corrected stellar mass, the fifth one corresponds to the Kroupa IMF stellar mass and the last one corresponds to the Kroupa IMF-corrected stellar mass. The IMF corrections to the stellar mass were made using equation 7.

The distribution of the stellar mass with respect to the virial mass may be analysed from two different points of view.

Case 1) Global behaviour. In this case we refer to the masses of galaxies as a set of dots to which a linear regression can be applied. This regression will give us the behaviour of the virial mass respect to the stellar mass, and this will give us information as to the possible existence of dark matter inside ETGs.

Case 2) Individual behaviour. In this case we consider the behaviour of the stellar mass with respect to the virial mass, one galaxy at a time. A relatively practical way to analyse this behaviour is to compare the intrinsic dispersion of luminous mass at a quasi-constant value of virial mass and viceversa. The analysis of the intrinsic dispersion might help us find the physical origin of this dispersion. It could also help us, although in a less direct fashion than in case 1, to investigate the presence of dark matter inside ETGs.

In what follows when we express the idea of quasi-constant mass, we shall mean mass intervals in the logarithm of width equal to 0.1.

In Appendix A we study in detail the distribution of virial mass vs. stellar mass and perform numerical simulations to analyse the possible presence of dark matter inside ETGs. The main results found are the following:

\begin{itemize}

\item The intrinsic dispersion of the virial mass vs. stellar mass of the different samples is at least three times larger that the mean mass error. This means that the extra dispersion might be due to dependencies of the virial and/or stellar mass on variables such as: redshift, wavelength and environment.

\item  The intrinsic dispersion of the virial mass vs. stellar mass depends on the mass of galaxies (see Figure 1 and section 7.3).

\item The geometry of the mass distribution of galaxies depends both on the intrinsic properties of the galaxies, as well as on the biases introduced by making arbitrary cuts to the samples. The effect of the different biases introduced is that the values of the slope obtained from the linear fits depend on the mass distribution (geometric effect), that is to say, the fit parameters depend on intrinsic properties of the galaxies and observational biases.

\item Due to the geometric effect, the set of calculated linear fit slopes to the virial vs. stellar masses is not an adequate set of parameters to perform analysis of intrinsic properties of the galaxies. This method is, therefore, not useful for investigating the presence of dark matter inside ETGs.

\item The weighted bisector fit applied to the mean value of the distribution at quasi-constant mass ($WBQ$ fit) is an adequate method for studying the global properties of the ETGs (for details see section 7.3 of Appendix A). This method may only be used as a first approximation to the study of dark matter inside ETGs given that the amount of dark matter might depend on variables such as mass (virial and stellar), wavelength, redshift and/or environment.

\item The intrinsic dispersion at quasi-constant mass does not depend on the geometric effect. This suggests that the intrinsic dispersion at quasi-constant mass might be a good tool for analysing the intrinsic properties of galaxies as functions of other variables, such as wavelength, redshift or environment.

\end{itemize}


\begin{figure*}
   \begin{center}



      \includegraphics[width=17cm]{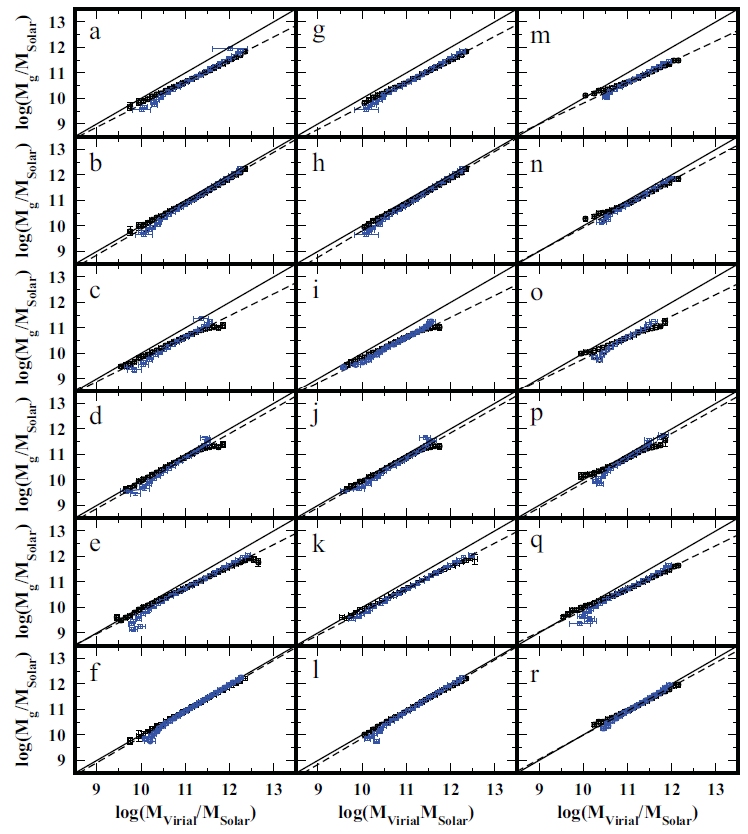}
      \caption{Distribution of the mean values of the virial and stellar mass from the Total (column 1), Morphologic (column 2) and Homogeneous (column 3) samples. The first row corresponds to the Salpeter IMF stellar mass, the second one corresponds to the Salpeter IMF-corrected stellar mass, the third one corresponds to the S\'ersic Salpeter stellar mass, the fourth one corresponds to the S\'ersic Salpeter IMF-corrected stellar mass, the fifth one corresponds to the Kroupa IMF stellar mass and the last one corresponds to the Kroupa IMF-corrected stellar mass. The black dots represent the mean values of the stellar mass at quasi-constant virial mass and the blue squares represent the mean values of the virial mass at quasi-constant stellar mass. The dotted line corresponds to the $WBQ$ fit. The solid line is the one-to-one line.}
         \label{FigVibStab}
         \end{center}
   \end{figure*}

\section{Dark matter inside the ETGs from the DR9}

In the previous section and appendix A we saw that the $WBQ$ fit is a good method to study the global properties of galaxies, particularly to investigate the presence of dark matter inside ETGs. We shall now present the results of applying this method to our 98,000 ETGs, as well as the application to our 19,000 ETGs homogeneous sample and also to the 27,000 ETGs morphological sample. The final errors have been obtained considering the stellar and virial mass errors as well as the error due to the $WBQ$ fit to the different samples. Here it is important to note that the $WBQ$ fit has been applied only to the mass interval where the homogeneous sample is complete, that is to say, for $log({\bf M_{virial}}) >$ 10.5 ($M_{g} \leq -20.0$).


In Figure 2 we show a mosaic of the behaviour of the stellar mass with respect to the virial mass for the ETGs Total, Morphologic and Homogeneous samples. The first row corresponds to the Salpeter IMF stellar mass, the second one corresponds to the Salpeter IMF-corrected stellar mass, the third one corresponds to the S\'ersic Salpeter stellar mass, the fourth one corresponds to the S\'ersic Salpeter IMF-corrected stellar mass, the fifth one corresponds to the Kroupa IMF stellar mass and the  last one corresponds to the Kroupa IMF-corrected stellar mass. Each graph shows the mean values of the luminous mass distribution at quasi-constant virial mass (black dots), the mean values of the virial mass distribution at quasi-constant stellar mass (blue squares) and the $WBQ$ fit (dotted line) to both point distributions. The solid line is the one-to-one line.

In Table 1 we show the equations of the fits and the estimations of dark matter inside galaxies of the different samples. The amount of dark matter was calculated as follows:

\begin{itemize}

\item We define 300 values for the logarithm of the virial mass (homogeneously distributed in the 9.5 - 12.5 interval)

\item For each value of the logarithm  of the virial mass we obtained the logarithm of the stellar mass using the following equation:

$log({\bf M_{g}}/{\bf M_{\odot}}) =  A\; log({\bf M_{virial}}/{\bf M_{\odot}}) + B$

where A and B represent the slope and the zero point obtained from the fits to the different samples (see equations in Table 1).

\item We calculate the percentage difference between the virial and stellar masses described previously.


\item We calculate the minimum, maximum and mean values of these differences.


\item We transform minimum, maximum and mean values to a linear scale. These quantities correspond to the estimated minimum, maximum and mean dark matter percentage for the galaxies in each one of the samples (see Table 1).

At this point it is important to point out that the errors associated with the different calculated parameters in this paper are obtained using the rules of error propagation in the following way:

	-Addition in quadrature if data are independent. This is the case for the slope and  zero point  errors of the fits to the different samples considered (see Section 3.3).

	-Straight addition if the data are not independent. This is the case for the errors in the dark matter estimates for the different samples and combinations thereof.

\end{itemize}

According to the information in Table 1 we find:

\begin{itemize}

\item The average amount of dark matter considering the mean values and only those samples with a universal IMF is 7.1\% $\pm$ 22.2\%. If we correct for a non universal IMF the amount of dark matter is  2.6\% $\pm$ 21.5\%. This result seems to indicate that correction for a non universal IMF reduces the estimated amount of dark matter (see discussion of this result in the following paragraphs).
	
\item The average amount of dark matter considering the mean values and only those samples with a universal IMF where the masses were obtained using de Vaucouleurs profiles is 7.7\% $\pm$ 21.7\%, whereas if we consider S\'ersic profiles the amount of dark matter is 7.9\% $\pm$ 23.7\%. Considering only those samples corrected for a non universal IMF where the masses were obtained using de Vaucouleurs profiles the amount of dark matter is 2.9\% $\pm$ 22.1\%, whereas if we consider S\'ersic profiles the amount of dark matter is 3.1\% $\pm$ 22.5\%. This suggest that the estimated amount of dark matter is not affected by the profile used in the calculation of luminous and virial mass.
	
\item The average amount of dark matter if we consider the mean values and only samples with a universal IMF in large intervals of redshift (total and morphological samples) is 7.1\% $\pm$ 22.2\%, whereas if we consider only the sample restricted in redshift (homogeneous sample) the amount of dark matter is 6.9\% $\pm$ 22.3\%. If we consider only samples corrected for non universal IMF in large intervals of redshift (total and morphological samples) the amount of dark matter is 2.7\% $\pm$ 21.3\%, whereas if we consider only the sample restricted in redshift (homogeneous sample) the amount of dark matter is 2.4\% $\pm$ 21.9\%. This suggests that the estimated amount of dark matter is similar when we move from a wide to a narrow redshift interval.

\item The average differences of dark matter between the maximum and minimum values for the non IMF corrected samples is 6.524. This relatively high difference as well as the equations corresponding to Table 1 suggest that there is a dependence of the dark matter on virial and stellar mass.

\item The average differences of dark matter between the maximum and minumum values for the IMF corrected samples is 1.823. This relatively small difference as well as the equations corresponding to Table 1 indicate that the amount of dark matter is approximately equal for the entire range of virial and stellar mass.


\end{itemize}

 As seen in Section 3, the correction for a non universal IMF \citep{dut13} has been obtained from models that consider the observed velocity dispersion of the galaxies in a sample. Our estimation of the virial mass was also obtained using the observed velocity dispersion, therefore the corrected stellar mass and the virial mass could be correlated. Besides the Dutton correction for non universal IMF assumes that the mass follows the light, that is, that the dark matter follows the same density profile as the stellar component, which, according to some authors \citep{koopmans2006,thomas2011}, is not valid for massive galaxies (they seem to be more affected by the correction, as can be seen in Figure 2, they are those that originally are further away from the line of slope 1) and therefore, that the estimation of the dark matter amount for corrected samples as well as its behaviour as a function of virial or stellar mass is not trustworthy.

Taking into consideration what we stated in the previous paragraph  the estimate we shall give in this paper will correspond to the average of the dark matter mean values from all the non corrected IMF samples, that is approximately 7\% $\pm$ 22\% of dark matter.

It is interesting to note that the difference between the mean value of dark matter for the IMF corrected and non corrected samples is relatively small and always contained within the errors.

As discussed in the introduction, the approximation to the dark matter problem within the effective radius presents different aspects. Some of the more recent and conspicuous studies are the following: \cite{gerhard2001}, based on a uniform dynamical analysis of the line-profile shapes of 21 mostly luminous, slowly rotating, and nearly round elliptical galaxies, find 10-40\% of dark matter. \cite{cap06} construct two-integral Jeans and three-integral Schwarzschild dynamical models for a sample of 25 E/S0 galaxies with SAURON integral-field stellar kinematics and find an average of 30\% of dark matter. \cite {thomas2007} survey axisymmetric Schwarzschild models for a sample of 17 Coma early-type galaxies and find 10-50\% of dark matter. \cite{williams2009} present comparisons between a large sample of dynamically determined stellar mass-to-light ratios and the predictions of stellar population models with which they find 15\% of dark matter. \cite{bar11} using gravitational lensing and stellar kinematics of the sixteen early-type lens galaxies from the SLACS Survey, at z = 0.08 - 0.33 find an average of 12\% of dark matter. However, if they use stellar population synthesis models with a Salpeter IMF, they find an average 31\% of dark matter, while if they use a Chabrier IMF the average dark matter found is 61\%.  \cite{thomas2011} compared dynamically derived stellar mass-to-light ratios with completely independent results from simple stellar population models from Coma early-type galaxies and find 28\%. \cite{nig11} comparing luminous mass (obtained from colours and a Salpeter IMF) and dynamical mass of approximately 90 000 ETGs find an almost negligible amount of dark matter. Finally, \cite{cap13}, using detailed axisymmetric dynamical models and deriving accurate total mass-to-light ratios (M/L) from a volume-limited and nearly mass-selected $ATLAS^{3D}$ sample of 260 early-type galaxies find 13\% of dark matter.

From the previous data we realise that the minimum estimate of dark matter is approximately 10\% and that the average of these estimates is closer to 30\%. On the other hand, the methods utilised by the different authors to estimate the amount of dark matter are varied and never coincide with the method used in this paper, so it is not possible to perform a direct comparison with our result. Besides, given the nature of our data and the procedure followed for the calculation of the virial and stellar masses, our result has an associated error which is relatively large, so that it may only be taken as a general tendency and the comparisons with the literature may only be qualitative. So, we may only affirm that the 7\% of dark matter estimated by us is smaller than the smallest value found in the literature (10\%) but this difference is within the associated error. On the other hand, our estimate is approximately four times smaller than the literature average (30\%). However, if we consider the associated error, our estimate is of the order of the literature average.

As discussed above, using a correction for a non universal IMF reduces the estimated amount of dark matter. However, these results cannot be taken as conclusive because the differences are within the associated error and the correction may not be precisely accurate. On the other hand, using linear fits to samples with high intrinsic dispersion (dispersion that appears to have a physical origin since its value is at least three times larger than the associated errors) is not appropriate to investigate the intrinsic properties of galaxies. The fitting method applied (using bins instead of points, see Appendix A) reduces the associated biases, however it is only a first approximation given that the intrinsic dispersion depends on the mass and it could also depend on other variables such as the wavelength, the environment and/or the redshift. These dependencies may also be correlated with the amount of dark matter and, to control them we must perform individual comparisons of the galaxy masses as functions of the different variables mentioned above. Another method to investigate these dependencies would be that proposed in Appendix A (section 7.3), that is to say, comparing the intrinsic dispersion at quasiconstant mass (virial and stellar) at different wavelengths, environment and redshift. In a forthcoming paper we shall discuss in depth these methods and we shall apply them to our galaxy samples.

\begin{table*}

\renewcommand{\footnoterule}{}  

\caption{Equations from the $WBQ$ fit to the different ETGs samples (see Figure 2) and dark matter estimations.}

\scalebox{0.35}{
\begin{tabular}{|l|c|c|c|c|c|}

\hline
\hline
{\Huge Name of the sample} & {\Huge Equation of the fit} & {\Huge} &{\Huge Amount of dark matter} &{\Huge} \\
\hline
\hline
& & & &\\
&  &{\Huge Minimum} &{\Huge Maximum} & {\Huge Mean}\\
& {\Huge Total samples} &  & & \\
\hline
 {\Huge de Vaucouleurs Salpeter-IMF stellar mass} & {\Huge $log({\bf M_{g}}/{\bf M_{\odot}}) =  (0.885 \pm 0.085)\; log({\bf M_{virial}}/{\bf M_{\odot}}) + (0.905 \pm 0.093)$ }  &{\Huge 4.545 $\pm$ 21.83} &{\Huge 9.809 $\pm$ 21.29} & {\Huge 7.397 $\pm$ 21.53} &{\Huge a} \\

{\Huge de Vaucouleurs Salpeter-IMF corrected stellar mass} & {\Huge $log({\bf M_{g}}/{\bf M_{\odot}}) =  (1.007 \pm 0.085)\; log({\bf M_{virial}}/{\bf M_{\odot}}) - (0.206 \pm 0.095)$  }  &{\Huge 2.183 $\pm$ 19.57} &{\Huge 3.381 $\pm$ 21.87} & {\Huge 2.732 $\pm$ 21.56} &{\Huge b}\\

{\Huge S\'ersic Salpeter-IMF stellar mass} & {\Huge $log({\bf M_{g}}/{\bf M_{\odot}}) =  (0. 861\pm 0.086)\; log({\bf M_{virial}}/{\bf M_{\odot}}) + (1.140 \pm 0.133)$  }  &{\Huge 4.375 $\pm$ 23.03} &{\Huge 11.01 $\pm$ 22.25} & {\Huge 7.968 $\pm$ 22.61} &{\Huge c}\\

{\Huge S\'ersic Salpeter-IMF corrected stellar mass} & {\Huge $log({\bf M_{g}}/{\bf M_{\odot}}) =  (0.980 \pm 0.086)\; log({\bf M_{virial}}/{\bf M_{\odot}}) + (0.065 \pm 0.098)$  }  &{\Huge 3.029 $\pm$ 22.08} &{\Huge 3.408 $\pm$ 21.70} & {\Huge 3.235 $\pm$ 19.80} &{\Huge d}\\

{\Huge Kroupa-IMF stellar mass} & {\Huge $log({\bf M_{g}}/{\bf M_{\odot}}) =  (0.874 \pm 0.083)\; log({\bf M_{virial}}/{\bf M_{\odot}}) + (1.101 \pm 0.084)$  }  &{\Huge 2.327 $\pm$ 21.15} &{\Huge 8.731 $\pm$ 20.66} & {\Huge 5.797 $\pm$ 20.88} &{\Huge e}\\

{\Huge Kroupa-IMF corrected stellar mass} & {\Huge $log({\bf M_{g}}/{\bf M_{\odot}}) =  (1.003 \pm 0.083)\; log({\bf M_{virial}}/{\bf M_{\odot}}) - (0.116 \pm 0.084)$  }  &{\Huge 1.446 $\pm$ 20.66} &{\Huge 2.121 $\pm$ 21.15} & {\Huge 1.756 $\pm$ 20.88} &{\Huge f}\\
\hline

& {\Huge Morphological samples} \\
\hline
 {\Huge de Vaucouleurs Salpeter-IMF stellar mass} & {\Huge $log({\bf M_{g}}/{\bf M_{\odot}}) =  (0.912 \pm 0.086)\; log({\bf M_{virial}}/{\bf M_{\odot}}) + (0.577 \pm 0.100)$  }  &{\Huge 6.278 $\pm$ 22.23} &{\Huge 9.634 $\pm$ 21.64} & {\Huge 8.096 $\pm$ 21.91} &{\Huge g}\\

{\Huge de Vaucouleurs Salpeter-IMF corrected stellar mass} & {\Huge $log({\bf M_{g}}/{\bf M_{\odot}}) =  (1.037 \pm 0.087)\; log({\bf M_{virial}}/{\bf M_{\odot}}) - (0.564 \pm 0.103)$  }  &{\Huge 1.869 $\pm$ 21.93} &{\Huge 5.151 $\pm$ 22.53} & {\Huge 3.373 $\pm$ 22.20} &{\Huge h}\\

{\Huge S\'ersic Salpeter-IMF stellar mass} & {\Huge $log({\bf M_{g}}/{\bf M_{\odot}}) =  (0.858 \pm 0.091)\; log({\bf M_{virial}}/{\bf M_{\odot}}) + (1.184 \pm 0.192)$  } &{\Huge 4.058 $\pm$ 25.61} &{\Huge 10.95 $\pm$ 24.49}  & {\Huge 7.791 $\pm$ 25.00} &{\Huge i}\\

{\Huge S\'ersic Salpeter-IMF corrected stellar mass} & {\Huge $log({\bf M_{g}}/{\bf M_{\odot}}) =  (0.976 \pm 0.092)\; log({\bf M_{virial}}/{\bf M_{\odot}}) + (0.128 \pm 0.144)$  }  &{\Huge 2.424 $\pm$ 24.63} &{\Huge 3.168 $\pm$ 23.88} & {\Huge 2.827 $\pm$ 21.18} &{\Huge j}\\

{\Huge Kroupa-IMF stellar mass} & {\Huge $log({\bf M_{g}}/{\bf M_{\odot}}) =  (0.902 \pm 0.084)\; log({\bf M_{virial}}/{\bf M_{\odot}}) + (0.794 \pm 0.089)$  }  &{\Huge 3.321 $\pm$ 21.49} &{\Huge 7.939 $\pm$ 20.98} & {\Huge 5.823 $\pm$ 21.22} &{\Huge k}\\

{\Huge Kroupa-IMF corrected stellar mass} & {\Huge $log({\bf M_{g}}/{\bf M_{\odot}}) =  (1.026 \pm 0.087)\; log({\bf M_{virial}}/{\bf M_{\odot}}) - (0.393 \pm 0.101)$  }  &{\Huge 1.253 $\pm$ 21.89} &{\Huge 3.539 $\pm$ 22.48} & {\Huge 2.300 $\pm$ 22.16} &{\Huge l}\\
\hline

& {\Huge Homogeneous samples} \\
\hline
 {\Huge de Vaucouleurs Salpeter-IMF stellar mass} & {\Huge $log({\bf M_{g}}/{\bf M_{\odot}}) =  (0.815 \pm 0.086)\; log({\bf M_{virial}}/{\bf M_{\odot}}) + (1.658 \pm 0.100)$  }  &{\Huge 2.412 $\pm$ 22.23} &{\Huge 12.06 $\pm$ 21.64} & {\Huge 7.637 $\pm$ 21.91 } &{\Huge m}\\

{\Huge de Vaucouleurs Salpeter-IMF corrected stellar mass} & {\Huge $log({\bf M_{g}}/{\bf M_{\odot}}) =  (0.925 \pm 0.088)\; log({\bf M_{virial}}/{\bf M_{\odot}}) + (0.683 \pm 0.113)$  }  &{\Huge 0.715 $\pm$ 23.00} &{\Huge 4.688 $\pm$ 22.34} & {\Huge 2.868 $\pm$ 22.65} &{\Huge n}\\

{\Huge S\'ersic Salpeter-IMF stellar mass} & {\Huge $log({\bf M_{g}}/{\bf M_{\odot}}) =  (0.845 \pm 0.088)\; log({\bf M_{virial}}/{\bf M_{\odot}}) + (1.324 \pm 0.154)$  }  &{\Huge 3.659 $\pm$ 23.99} &{\Huge 11.36 $\pm$ 23.09} & {\Huge 7.831 $\pm$ 23.51} &{\Huge o}\\

{\Huge S\'ersic Salpeter-IMF corrected stellar mass} & {\Huge $log({\bf M_{g}}/{\bf M_{\odot}}) =  (0.972 \pm 0.089)\; log({\bf M_{virial}}/{\bf M_{\odot}}) + (0.155 \pm 0.114)$  }  &{\Huge 2.690 $\pm$ 23.26} &{\Huge 3.592 $\pm$ 22.59} & {\Huge 3.179 $\pm$ 20.49} &{\Huge p}\\

{\Huge Kroupa-IMF stellar mass} & {\Huge $log({\bf M_{g}}/{\bf M_{\odot}}) =  (0.847 \pm 0.085)\; log({\bf M_{virial}}/{\bf M_{\odot}}) + (1.410 \pm 0.095)$  }  &{\Huge 1.054 $\pm$ 21.87} &{\Huge 9.256 $\pm$ 21.32} & {\Huge 5.498 $\pm$ 21.57} &{\Huge q}\\

{\Huge Kroupa IMF corrected stellar mass} & {\Huge $log({\bf M_{g}}/{\bf M_{\odot}}) =  (0.948 \pm 0.088)\; log({\bf M_{virial}}/{\bf M_{\odot}}) + (0.511 \pm 0.110)$  }  &{\Huge 0.412 $\pm$ 22.93} &{\Huge 2.560 $\pm$ 22.29} & {\Huge 1.199 $\pm$ 22.58} &{\Huge r}\\
\hline

\end{tabular}}
\end{table*}

\section{Conclusions}

The analysis of the distribution of stellar mass with respect to virial mass on several samples of ETGs from the DR9 has yielded the following results:

\begin{enumerate}

\item The distribution of the virial mass vs. stellar mass has a relatively high intrinsic dispersion. This dispersion might have a physical origin given that it is at least three times larger that the associated errors.

\item The values of the parameters obtained from linear regressions performed on the mass distribution ($BCES_{Bis}$ method) are affected by a geometric effect (see Appendix A). This method is, therefore, not useful for investigating the intrinsic properties of the ETGs.

\item Monte Carlo simulations (see Appendix A) proved that the weighted bisector fit applied to the mean value of the distribution at quasi-constant mass ($WBQ$ fit) is an adequate method for studying the global properties of the ETGs (for more details see section 7.3). This method may only be used as a first approximation to the study of dark matter inside ETGs given that the amount of dark matter might depend on variables such as mass (virial and stellar), wavelength, redshift and/or environment.

\item Application of the $WBQ$ fit (see Section 5 and Appendix A) to the different samples produces the following results:

 \begin{itemize}

\item The amount of dark matter (mean value) considering only those samples with a universal IMF is 7.1\% $\pm$ 22.2\%. If we correct for a non universal IMF the amount of dark matter is  2.6\% $\pm$ 21.5\%. This result seems to indicate that correction for a non universal IMF reduces the estimated amount of dark matter.

\item The amount of dark matter (mean value) considering only those samples where the masses were obtained using de Vaucouleurs profiles is similar to that where we only consider S\'ersic profiles. This suggests that the estimated amount of dark matter is not affected by the profile used in the calculation of luminous and virial mass.

\item The amount of dark matter (mean value) if we consider only samples in large intervals of redshift (total and morphological samples) is similar to that when we consider only the sample restricted in redshift (homogeneous sample). This suggests that the estimated amount of dark matter does not change when we move from a wide to a narrow redshift interval.

\item The average differences of dark matter (approximately 7\%) between the maximum and minumum values for the non IMF corrected samples and the corresponding equations shown in Table 1 suggest a dependence of the amount of dark matter on virial and stellar mass (see section 5).

\item The average differences of dark matter (smaller than 2\%) between the maximum and minumum values for the non IMF corrected samples and the corresponding equations shown in Table 1 suggest that the amount of dark matter is approximately equal for the entire range of virial and stellar mass (see Section 5).

\end{itemize}

The previous results suggest that a correction for a non universal IMF not only reduces the estimated amount of dark matter but also the dependence of the amount of dark matter on virial and stellar mass. However these results cannot be considered conclusive because the applied correction may not be precisely adequate (see Section 5). Considering this fact, the estimate of dark matter that we give in this paper corresponds to the average of the non IMF corrected samples, that is approximately 7\% $\pm$ 22\% of dark matter.


\item The amount of dark matter found inside ETGs in relevant works in the literature varies between 10\% and 60\%, with an average value of approximately 30\% (see Section 5). Our estimate of the amount of dark matter is of the order of the smallest value given in the literature and four times lower than the average (30\%) of literature estimates. However, given the variety of methods for the calculation of the amount of dark matter used in the literature and that none of them is similar to the one used in this paper, and also that the values of our errors are relatively large, our results may only be taken as a general tendency. In a forthcoming paper we shall use other methods to estimate the amount of dark matter (see methods proposed in section 7.3 of Appendix A), and we shall also analyse the behaviour of the amount of dark matter as function of mass, wavelength, environment and redshift.

\end{enumerate}

\section*{Acknowledgments}

We would like to dedicate this humble work to the memory of Mrs. Eutiquia Netro Castillo, an extraordinary woman.

We would like to thank Instituto de Astronom\'{\i}a y Meteorolog\'{\i}a (UdG, M\'exico), Instituto de Astronom\'{\i}a (UNAM, M\'exico), and Instituto de Astrof\'{\i}sica de Andaluc\'{\i}a (IAA, Espa\~na) for all the facilities provided for the realisation of this project. Our thanks are due to Luis M. Borruel and Isaac A. Perez for efficient computational assistance. Alberto Nigoche-Netro and G. Ramos-Larios acknowledge support from CONACyT and PROMEP (M\'exico). Patricio Lagos is supported by a PostDoctoral Grant SFRH/BPD/72308/2010, funded by FCT (Portugal) and Funda\c{c}\~{a}o para a Ci\^{e}ncia e a Tecnologia (FCT) under project FCOMP-01-0124-FEDER-029170 (Reference FCT PTDC/FIS-AST/3214/2012), funded by the FEDER program. We express our deepest appreciation to the anonymous referee whose comments and suggestions greatly improved the presentation of this paper.



\section{Appendix A. Distribution of the stellar mass as a function of the virial mass of ETGs}

The distribution of the stellar mass with respect to the virial mass (Figure 1) may be analysed from two different points of view.

Case 1) Global behaviour. In this case we refer to the masses of galaxies as a set of dots to which a linear regression can be applied. This regression will give us the behaviour of the virial mass respect to the stellar mass, and this will give us information as to the possible existence of dark matter inside ETGs.

Case 2) Individual behaviour. In this case we consider the behaviour of the stellar mass with respect to the virial mass, one galaxy at a time. A relatively practical way to analyse this behaviour is to compare the intrinsic dispersion of luminous mass at a quasi-constant value of virial mass and viceversa. The analysis of the intrinsic dispersion might help us find the physical origin of this dispersion. It could also help us, although in a less direct fashion than in case 1, to investigate the presence of dark matter inside ETGs.

We must define in a rigorous way the intrinsic dispersion of a set of points in which X represents the independent variable and Y the dependent variable. Here it is interesting to distinguish two properties of the dot distribution, the first property refers to the width of the dot distribution around the variable Y and the second refers to the mean value of the Y coordinates. According to this, the intrinsic dispersion is defined as the width of the dot distribution around the mean value of the variable Y, which, in mathematical terms, would be the standard deviation of the dot distribution with respect to the mean value of the Y coordinate.

In the following sections we shall deal with cases 1 and 2 to analyse the possible presence of dark matter inside ETGs.

\subsection{Dark matter inside ETGs. Case 1}

This case needs hardly any explanation because all that is required is to perform a linear fit to the virial mass vs. the stellar mass. The slope of this fit would be related to the presence of dark matter inside ETGs. A slope with a value of one means that there is the same amount of stellar mass as  virial mass and, therefore, there is no need to invoke the presence of dark matter. An important point to comment here is the method followed to perform the fit used to obtain the values of the parameters of the linear regression, since it has been demonstrated that these parameters are very sensitive to several of the data properties, including: the errors in the variables, the error correlation, the data dispersion and using one or the other variable as the dependent variable of the fit. The {\it Bivariate correlated errors and intrinsic scatter bisector} ($BCES_{Bis}$) method \citep{iso90,akr96} takes into consideration the different sources of bias listed above, so this is the first method that we shall use to perform fits in this paper. It is important to mention here that, in this case, possible dependencies of the amount of dark matter on luminosity, redshift, wavelength or environment are not considered.

\subsection{The geometric effect in the parameters of the linear regressions performed on the observed distributions of virial vs. stellar mass.}

 As may be seen from Figure 1 the galaxies are distributed in a cloud near the one-to-one line (solid line). The intrinsic dispersion of the distribution of the sample with stellar mass obtained using the Salpeter IMF is, on average, 0.015 dex greater than Kroupa. In the case of the IMF-corrected and non-corrected samples the difference in the intrinsic dispersion is within the associated error, that is to say, the intrinsic dispersion does not diminish when the stellar masses are corrected by the non-universal IMF. It is important to say that the intrinsic dispersion, in all cases, is relatively high (approximately 0.2 dex) and that this dispersion is at least three times larger that the mean mass error. This means that the intrinsic dispersion might have an unknown physical component. The extra dispersion might be due to dependencies of the virial and/or stellar mass on variables such as: redshift, wavelength, luminosity and environment.

 In \cite{nig08,nig09,nig10} it has been demonstrated that for linear fits to sets of data with a high intrinsic dispersion, the values of the parameters of the linear fits performed to these points depend on the geometric form of the point distribution. This geometric form may depend on physical properties of the galaxies as well as on observational biases and arbitrary cuts performed on the observed samples. This effect has been called the geometric effect (\citealt{nig08}). In previous works (\citealt{nig08}; \citealt{nig09}; \citealt{nig10}) it has been established that the geometric effect does not allow the determination of the physical properties of the samples of ETGs under study. For the virial vs. stellar mass distribution we have a relatively high intrinsic dispersion so that the straight line must be affected by the said geometric effect. In the following section we shall quantify the geometric effect on the values of the parameters of the linear regressions performed to the virial and stellar mass distribution.

\subsubsection{Simulated distribution of virial vs. stellar mass}

To quantify the geometric effect in the linear regressions made to our galaxy samples, we have made Monte Carlo simulations in which we have
considered that galaxies have similar distribution on the M$_{virial}$ vs. M$_g$ plane as that observed for the real data. The simulations were performed with 100 000 dots. The position of these points were generated by a random number generator within specific limits, either inside a rectangular or ellipsoidal region. The number of points decreases smoothly beyond the limits of the rectangular or ellipsoidal regions in order to give the simulations a better likeness to the real distribution of galaxies on the plane under study.

\begin{itemize}

\item {\it Simulated distribution of virial vs. stellar mass considering constant intrinsic dispersion (rectangular simulation)}.

In Figure 3 we see the simulated galaxy distribution in which the intrinsic dispersion at different virial masses is the same. The value of this intrinsic dispersion is 0.2 dex and is approximately equal to the value of the maximum dispersion of the real distribution. The simulated distribution of galaxies has been taken around a straight line of slope equal to 1 that passes through the origin, that is to say, the global behaviour of the virial vs. stellar mass is the same (no dark matter) and the individual behaviour of each type of mass (virial, stellar) is affected only by the dispersion which is the same at any value of the mass. If there is any physical property that is responsible for the intrinsic dispersion, this property would be the same and contribute the same amount over all the mass range.


\begin{figure}
   \centering
   \includegraphics[width=9cm]{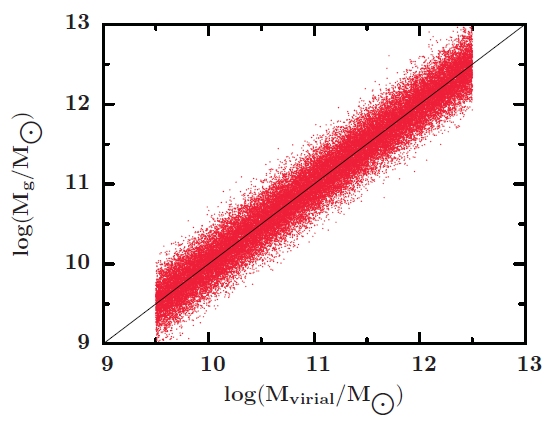}
      \caption{Distribution of the virial vs. stellar mass of the rectangular simulation. The solid line is the one-to-one line.}
         \label{FigVibStab}
   \end{figure}


\begin{figure}
   \centering
   \includegraphics[angle=0,width=8cm]{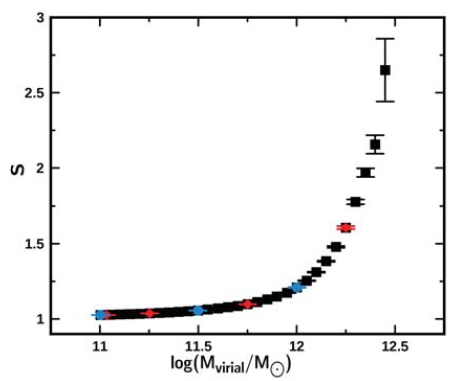}
      \caption{Behaviour of the $BCES_{Bis}$ fit slope in increasing virial mass intervals (lower mass cut-off) from the rectangular simulation. Each point corresponds to the mean value of the mass of the galaxies contained in each mass interval analysed (see Table 2). Circles, diamonds and squares represent intervals that progressively increase their width in amounts of 1.0, 0.5, and 0.1 in the abscissa.}
         \label{FigVibStab}
   \end{figure}

The analysis of the geometric effect on the linear fits to the mass distribution may be performed in two ways: increasing mass intervals and narrow mass intervals (for more details see \citealt{nig08}; \citealt{nig09}; \citealt{nig10}).

Table 2 presents the results of the analysis using increasing virial mass intervals (lower mass cut-off). Column 1 gives the interval of virial mass in which the galaxies are contained, column 2 indicates the number of galaxies in this interval, column 3 presents the value of the slope obtained from the linear fit to the data, column 4 gives the zero point and finally, column 5 indicates the intrinsic dispersion found from the data.


\begin{table*}

\begin{minipage}[t]{\columnwidth}

\renewcommand{\footnoterule}{}  

\caption{Coefficients of the $BCES_{Bis}$ fit for the rectangular simulation of galaxies in increasing-mass-intervals (lower mass cut-off).}

\vskip1.0cm

\begin{tabular}{cccccc}

\hline
$MI$   \footnote{Mass interval within which the galaxies are distributed.}         & $N$  \footnote{Number of galaxies in the mass interval.} & $S$ \footnote{Slope of the $BCES_{Bis}$ fit .}              & $A$   \footnote{Zero point of the $BCES_{Bis}$ fit.}     & $\sigma_{mass}$ \footnote{Intrinsic dispersion of the distribution.}\\

\hline


\multicolumn{5}{c}{}\\
\multicolumn{5}{c}{Intervals that progressively increase their width in amount of 0.1}\\
\multicolumn{5}{c}{}\\

$12.4 \geq M > 12.5$ & $3.299 \times 10^{3}$   & 2.6500 $\pm$ 0.2090 & -20.547 $\pm$ 2.6040 & 0.207 \\
$12.3 \geq M > 12.5$ & $6.714 \times 10^{3}$  & 2.1560 $\pm$ 0.0610 & -14.338  $\pm$ 0.7610 & 0.209 \\
$12.2 \geq M > 12.5$ & $1.005 \times 10^{4}$ & 1.9690 $\pm$ 0.0280 & -11.966  $\pm$ 0.3480 & 0.218 \\
$12.1 \geq M > 12.5$ & $1.342 \times 10^{4}$ & 1.7780 $\pm$ 0.0160 & -9.5710  $\pm$ 0.1920 & 0.232 \\
$12.0 \geq M > 12.5$ & $1.677 \times 10^{4}$ & 1.6060 $\pm$ $9.8150 \times 10^{-3}$ & -7.4230 $\pm$ 0.1200 & 0.247 \\
$11.9 \geq M > 12.5$ & $2.010 \times 10^{4}$ & 1.4790 $\pm$ $6.7370 \times 10^{-3}$ & -5.8390 $\pm$ 0.0820 & 0.264 \\
$11.8 \geq M > 12.5$ & $2.346 \times 10^{4}$ & 1.3840 $\pm$ $4.7670 \times 10^{-3}$ & -4.6720  $\pm$ 0.0580 & 0.285 \\
$11.7 \geq M > 12.5$ & $2.678 \times 10^{4}$ & 1.3110 $\pm$ $3.5490 \times 10^{-3}$ & -3.7600 $\pm$ 0.0430 & 0.305 \\
$11.6 \geq M > 12.5$ & $3.007 \times 10^{4}$ & 1.2540 $\pm$ $2.7350 \times 10^{-3}$ & -3.0560 $\pm$ 0.0330 & 0.327 \\
$11.5 \geq M > 12.5$ & $3.338 \times 10^{4}$ & 1.2090 $\pm$ $2.1380 \times 10^{-3}$ & -2.5050  $\pm$ 0.0260 & 0.35 \\
$11.4 \geq M > 12.5$ & $3.673 \times 10^{4}$ & 1.1740 $\pm$ $1.7170 \times 10^{-3}$ & -2.0840  $\pm$ 0.0210 & 0.374 \\
$11.3 \geq M > 12.5$ & $4.017 \times 10^{4}$ & 1.1490 $\pm$ $1.3790 \times 10^{-3}$ & -1.7730  $\pm$ 0.0160 & 0.399 \\
$11.2 \geq M > 12.5$ & $4.344 \times 10^{4}$ & 1.1300 $\pm$ $1.1480 \times 10^{-3}$ & -1.5380  $\pm$ 0.0140 & 0.424 \\
$11.1 \geq M > 12.5$ & $4.680 \times 10^{4}$ & 1.1120 $\pm$ $9.5710 \times 10^{-4}$ & -1.3240  $\pm$ 0.0110 & 0.450 \\
$11.0 \geq M > 12.5$ & $5.012 \times 10^{4}$ & 1.0980 $\pm$ $8.1080 \times 10^{-4}$ & -1.1540  $\pm$ $9.5700 \times 10^{-3}$ & 0.476 \\
$10.9 \geq M > 12.5$ & $5.345 \times 10^{4}$ & 1.0860 $\pm$ $6.9180 \times 10^{-4}$ & -1.0100  $\pm$ $8.1430 \times 10^{-3}$ & 0.502 \\
$10.8 \geq M > 12.5$ & $5.679 \times 10^{4}$ & 1.0780$\pm$ $5.9540 \times 10^{-4}$ & -0.9050  $\pm$ $6.9890 \times 10^{-3}$ & 0.529 \\
$10.7 \geq M > 12.5$ & $6.015 \times 10^{4}$ & 1.0700 $\pm$ $5.1590 \times 10^{-4}$ & -0.8120  $\pm$ $6.0430 \times 10^{-3}$ & 0.556 \\
$10.6 \geq M > 12.5$ & $6.366 \times 10^{4}$ & 1.0630 $\pm$ $4.4810 \times 10^{-4}$ & -0.7210  $\pm$ $5.2380 \times 10^{-3}$ & 0.585 \\
$10.5 \geq M > 12.5$ & $6.696 \times 10^{4}$ & 1.0560 $\pm$ $3.9570 \times 10^{-4}$ & -0.6480  $\pm$ $4.6190 \times 10^{-3}$ & 0.611 \\
$10.4 \geq M > 12.5$ & $7.020 \times 10^{4}$ & 1.0510 $\pm$ $3.5130 \times 10^{-4}$ & -0.5870  $\pm$ $4.0960 \times 10^{-3}$ & 0.637 \\
$10.3 \geq M > 12.5$ & $7.348 \times 10^{4}$ & 1.0470 $\pm$ $3.1310 \times 10^{-4}$ & -0.5370  $\pm$ $3.6480 \times 10^{-3}$ & 0.665 \\
$10.2 \geq M > 12.5$ & $7.675 \times 10^{4}$ & 1.0430 $\pm$ $2.8060 \times 10^{-4}$ & -0.4900  $\pm$ $3.2690 \times 10^{-3}$ & 0.692 \\
$10.1 \geq M > 12.5$ & $8.005 \times 10^{4}$ & 1.0400 $\pm$ $2.5240 \times 10^{-4}$ & -0.4480 $\pm$ $2.9430 \times 10^{-3}$ & 0.719 \\
$10.0 \geq M > 12.5$ & $8.335 \times 10^{4}$ & 1.0370 $\pm$ $2.2770 \times 10^{-4}$ & -0.4150 $\pm$ $2.6570 \times 10^{-3}$ & 0.747 \\
$9.9 \geq M > 12.5$ & $8.665  \times 10^{4}$ & 1.0340 $\pm$ $2.0610 \times 10^{-4}$ & -0.3830 $\pm$ $2.4090 \times 10^{-3}$ & 0.775 \\
$9.8 \geq M > 12.5$ & $9.005 \times 10^{4}$ & 1.0320 $\pm$ $1.8670 \times 10^{-4}$ & -0.3520 $\pm$ $2.1860 \times 10^{-3}$ & 0.804 \\
$9.7 \geq M > 12.5$ & $9.347 \times 10^{4}$ & 1.0300 $\pm$ $1.6980 \times 10^{-4}$ & -0.3300 $\pm$ $1.9960 \times 10^{-3}$ & 0.833 \\
$9.6 \geq M > 12.5$ & $9.648 \times 10^{4}$ & 1.0280 $\pm$ $1.5510 \times 10^{-4}$ & -0.3050 $\pm$ $1.8310 \times 10^{-3}$ & 0.861 \\
$9.5 \geq M > 12.5$ & $1.000 \times 10^{5}$ & 1.0260 $\pm$ $1.4290 \times 10^{-4}$ & -0.2479 $\pm$ $1.6950 \times 10^{-3}$ & 0.888 \\

\multicolumn{5}{c}{}\\

\multicolumn{5}{c}{Intervals that progressively increase their width in amount of 0.5}\\

\multicolumn{5}{c}{}\\

$12.0 \geq M > 12.5$ & $1.667 \times 10^{4}$   & 1.6060 $\pm$ $9.8150 \times 10^{-3}$ & -7.4230 $\pm$ 0.1200 & 0.247 \\
$11.5 \geq M > 12.5$ & $3.338 \times 10^{4}$   & 1.2090 $\pm$ $2.1380 \times 10^{-3}$ & -2.5050 $\pm$ 0.0260 & 0.350 \\
$11.0 \geq M > 12.5$ & $5.012 \times 10^{4}$   & 1.0980 $\pm$ $8.1080 \times 10^{-4}$ & -1.1540 $\pm$ $9.5700 \times 10^{-3}$ & 0.476 \\
$10.5 \geq M > 12.5$ & $6.696 \times 10^{4}$   & 1.0560 $\pm$ $3.9570 \times 10^{-4}$ & -648.00 $\pm$ $4.6190 \times 10^{-3}$ & 0.611 \\
$10.0 \geq M > 12.5$ & $8.335 \times 10^{4}$   & 1.0370 $\pm$ $2.2770 \times 10^{-4}$ & -0.4150 $\pm$ $2.6570 \times 10^{-3}$ & 0.747 \\
$9.5 \geq M > 12.5$ & $1.000 \times 10^{5}$   & 1.0260 $\pm$ $1.4290 \times 10^{-4}$ & -0.2840 $\pm$ $1.6950 \times 10^{-3}$ & 0.888 \\

\multicolumn{5}{c}{}\\

\multicolumn{5}{c}{Intervals that progressively increase their width in amount of 1.0} \\

\multicolumn{5}{c}{}\\

$11.5 \geq M > 12.5$  & $3.338 \times 10^{4}$  &  1.2090 $\pm$ $2.1380 \times 10^{-3}$ & -2.5050 $\pm$ 0.0260 & 0.349 \\
$10.5 \geq M > 12.5$  & $6.696 \times 10^{4}$  &  1.0560 $\pm$ $3.9570 \times 10^{-4}$ & -0.6480 $\pm$ $4.6190 \times 10^{-3}$ & 0.611 \\
$9.5 \geq M > 12.5$  & $1.000 \times 10^{5}$  &  1.0260 $\pm$ $1.4290 \times 10^{-4}$ & -0.2840 $\pm$ $1.6950 \times 10^{-3}$ & 0.888 \\

\hline

\end{tabular}
\end{minipage}
\end{table*}


\begin{table*}

\begin{minipage}[t]{\columnwidth}

\renewcommand{\footnoterule}{}  

\caption{Coefficients of the $BCES_{Bis}$ fit for the rectangular simulation of galaxies in narrow-mass-intervals.}

\vskip1.0cm

\begin{tabular}{cccccc}

\hline
$MI$   \footnote{Mass interval within which the galaxies are distributed.}         & $N$  \footnote{Number of galaxies in the mass interval.} & $S$ \footnote{Slope of the $BCES_{Bis}$ fit .}              & $A$   \footnote{Zero point of the $BCES_{Bis}$ fit.}     & $\sigma_{mass}$ \footnote{Intrinsic dispersion of the distribution.}\\

\hline

\multicolumn{5}{c}{}\\
\multicolumn{5}{c}{Intervals of width 0.1}\\
\multicolumn{5}{c}{}\\

$9.5 \geq M > 9.6$ & $3.1650 \times 10^{3}$   & 2.0220 $\pm$ 0.1950 & -9.7540 $\pm$ 1.8610 & 0.203 \\
$9.6 \geq M > 9.7$ & $3.3690 \times 10^{3}$  & 2.5660 $\pm$ 0.1950 & -15.109  $\pm$ 1.8860 & 0.201 \\
$9.7 \geq M > 9.8$ & $3.4160 \times 10^{3}$ & 2.3160 $\pm$ 0.1920 & -12.836  $\pm$ 1.8730 & 0.204 \\
$9.8 \geq M > 9.9$ & $3.4030 \times 10^{3}$ & 2.1700 $\pm$ 0.1880 & -11.521 $\pm$ 1.8530 & 0.198 \\
$9.9 \geq M > 10.0$ & $3.2960 \times 10^{3}$ & 2.0800 $\pm$ 0.1880 & -10.744  $\pm$ 1.8750 & 0.199 \\
$10.0 \geq M > 10.1$ & $3.3040 \times 10^{3}$ & 2.1150 $\pm$ 0.1950 & -11.209 $\pm$ 1.9560 & 0.200 \\
$10.1 \geq M > 10.2$ & $3.2990 \times 10^{3}$ & 2.2170 $\pm$ 0.1930 & -11.438 $\pm$ 1.0590 & 0.202 \\
$10.2 \geq M > 10.3$ & $3.2700 \times 10^{3}$ & 2.2080 $\pm$ 0.1950 & -12.382  $\pm$ 1.9990 & 0.201 \\
$10.3 \geq M > 10.4$ & $3.2800 \times 10^{3}$ & 2.3280 $\pm$ 0.1990 & -13.747  $\pm$ 2.0550 & 0.200 \\
$10.4 \geq M > 10.5$ & $3.2400 \times 10^{3}$ & 2.1500 $\pm$ 0.1910 & -12.013  $\pm$ 1.9930 & 0.200\\
$10.5 \geq M > 10.6$ & $3.3000 \times 10^{3}$ & 2.2720 $\pm$ 0.1970 & -13.418  $\pm$ 2.0830 & 0.204 \\
$10.6 \geq M > 10.7$ & $3.5080 \times 10^{3}$ & 2.3520 $\pm$ 0.1940 & -14.397  $\pm$ 2.0650 & 0.207 \\
$10.7 \geq M > 10.8$ & $3.3620 \times 10^{3}$ & 2.2940 $\pm$ 0.1930 & -13.913  $\pm$ 2.0750 & 0.202 \\
$10.8 \geq M > 11.9$ & $3.3350 \times 10^{3}$ & 2.5410 $\pm$ 0.2000 & -16.729  $\pm$ 2.1670 & 0.201 \\
$10.9 \geq M > 11.0$ & $3.3300 \times 10^{3}$ & 2.7610 $\pm$ 0.2010 & -19.278  $\pm$ 2.1980 & 0.204 \\
$11.0 \geq M > 11.1$ & $3.3200 \times 10^{3}$ & 2.4120 $\pm$ 0.1970 & -15.596  $\pm$ 2.1720 & 0.204 \\
$11.1 \geq M > 11.2$ & $3.3640 \times 10^{3}$ & 2.1140 $\pm$ 0.1890 & -12.417 $\pm$ 2.1060 & 0.201 \\
$11.2 \geq M > 11.3$ & $3.2680 \times 10^{3}$ & 2.1700 $\pm$ 0.2000 & -13.158  $\pm$ 2.2510 & 0.206 \\
$11.3 \geq M > 11.4$ & $3.4380 \times 10^{3}$ & 2.4310 $\pm$ 0.1850 & -16.239  $\pm$ 2.0940 & 0.199 \\
$11.4 \geq M > 11.5$ & $3.3540 \times 10^{3}$ & 2.1630 $\pm$ 0.1880 & -13.317  $\pm$ 2.1540 & 0.203 \\
$11.5 \geq M > 11.6$ & $3.3070 \times 10^{3}$ & 2.3130 $\pm$ 0.1900 & -15.165 $\pm$ 2.1950 & 0.199 \\
$11.6 \geq M > 11.7$ & $3.2940 \times 10^{3}$ & 2.3330 $\pm$ 0.1900 & -15.53 $\pm$ 2.2160 & 0.203 \\
$11.7 \geq M > 11.8$ & $3.3200 \times 10^{3}$ & 2.2610 $\pm$ 0.1970 & -14.813 $\pm$ 2.3140 & 0.200 \\
$11.8 \geq M > 11.9$ & $3.3570 \times 10^{3}$ & 2.5980 $\pm$ 0.1930 & -18.944 $\pm$ 2.2840 & 0.200 \\
$11.9 \geq M > 12.0$ & $3.3350 \times 10^{3}$ & 2.3560 $\pm$ 0.1980 & -16.202 $\pm$ 2.3670 & 0.204 \\
$12.0 \geq M > 12.1$ & $3.3490 \times 10^{3}$ & 2.5390 $\pm$ 0.1980 & -18.539 $\pm$ 2.3840 & 0.201 \\
$12.1 \geq M > 12.2$ & $3.3700 \times 10^{3}$ & 2.2590 $\pm$ 0.1950 & -15.296 $\pm$ 2.3650 & 0.201 \\
$12.2 \geq M > 12.3$ & $3.3340 \times 10^{3}$ & 2.3440 $\pm$ 0.1880 & -16.466 $\pm$ 2.2990 & 0.199 \\
$12.3 \geq M > 12.4$ & $3.4130 \times 10^{3}$ & 2.3790 $\pm$ 0.1900 & -17.031 $\pm$ 2.3510 & 0.200 \\
$12.4 \geq M > 12.5$ & $3.3010 \times 10^{3}$ & 2.6530 $\pm$ 0.2090 & -20.583 $\pm$ 2.6000 & 0.207 \\

\multicolumn{5}{c}{}\\
\multicolumn{5}{c}{Intervals of width 0.5}\\

\multicolumn{5}{c}{}\\

$9.5 \geq M > 10.0$ & $1.6650 \times 10^{4}$   & 1.6020 $\pm$ $9.9480 \times 10^{-3}$ & -5.8740 $\pm$ 0.0970 & 0.244 \\
$10.0 \geq M > 10.5$ & $1.6390 \times 10^{4}$   & 1.6030 $\pm$ $9.8230 \times 10^{-3}$ & -6.1810 $\pm$ 0.1010 & 0.246 \\
$10.5 \geq M > 11.0$ & $1.6480 \times 10^{4}$   & 1.6090 $\pm$ $9.9690 \times 10^{-3}$ & -6.5480 $\pm$ 0.1070 & 0.247 \\
$11.0 \geq M > 11.5$ & $1.6740 \times 10^{4}$   & 1.6020 $\pm$ $9.8830 \times 10^{-3}$ & -6.7710 $\pm$ 0.1110 & 0.247 \\
$11.5 \geq M > 12.0$ & $1.6610 \times 10^{4}$   & 1.5910 $\pm$ $9.8030 \times 10^{-3}$ & -6.9390 $\pm$ 0.1150 & 0.245 \\
$12.0 \geq M > 12.5$ & $1.6670 \times 10^{4}$   & 1.6060 $\pm$ $9.8150 \times 10^{-3}$ & -7.4230 $\pm$ 0.1200 & 0.247 \\


\multicolumn{5}{c}{}\\
\multicolumn{5}{c}{Intervals of width 1.0} \\

\multicolumn{5}{c}{}\\

$9.5 \geq M > 10.5$  & $3.3040 \times 10^{4}$  &  1.2120 $\pm$ $2.1450 \times 10^{-3}$ & -2.1200 $\pm$ 0.02100 & 0.349 \\
$10.5 \geq M > 11.5$  & $3.3580 \times 10^{4}$  &  1.2130 $\pm$ $2.1430 \times 10^{-3}$ & -2.3380 $\pm$ 0.02400 & 0.352 \\
$11.5 \geq M > 12.5$  & $3.3380 \times 10^{4}$  &  1.2090 $\pm$ $2.1380 \times 10^{-3}$ & -2.5050 $\pm$ 0.02600 & 0.350 \\


\hline

\end{tabular}
\end{minipage}
\end{table*}

In Figure 4 we present the behaviour of the slope of the linear fit as a function of the logarithm of the virial mass of the galaxies (increasing intervals). The plotted value of the logarithm of the virial mass corresponds to the mean value in each interval. In Figure 4 we can also see that, in spite of the fact that the mass distribution has the same intrinsic dispersion and that the galaxies are distributed around a straight line of slope 1 (galaxies have the same physical properties at any value of the virial mass), the fits depend on the interval of virial mass which is being considered. It may be appreciated that the value of the slope tends to 1 when the mass interval is relatively wide. The slope of the fit differs from the expected value ($m=1$) by less than 10\% when the virial mass interval in the logarithm is approximately equal to 1.5 or less.

In Table 3 we present the results of the analysis considering mass intervals of the same width at different values of the virial mass. The structure of Table 3 is the same as that of Table 2. In this table we see that the intrinsic dispersion approaches the expected value as the mass interval becomes narrower. If we calculate the intrinsic dispersion at constant mass the obtained value corresponds to the expected value.

In Figure 5 we present the behaviour of the slope of the linear fit as a function of the logarithm of the virial mass of the galaxies (same width intervals). The plotted value of the logarithm corresponds to the mean value in each interval of mass. This graph shows the slopes taken from Table 3. We show the slopes for intervals of width equal to 1 (blue dots), the slopes for intervals of width equal to 0.5 (red diamonds) and for intervals with width equal to 0.1 (black squares). We notice that for mass intervals of the same width the values of the obtained slopes are similar, although they differ from the expected value. The value of the slope approaches the expected value only when the mass interval is relatively wide.

\item {\it Simulated distribution of the virial mass vs. the stellar mass considering that the intrinsic dispersion is dependent on the virial and stellar mass (ellipsoidal simulation).}

Given that the real galaxy distribution does not have the rectangular form analysed above, it is important to study the behaviour of a simulated sample that has, as close as possible, the characteristics of the real sample, that is to say, that the intrinsic dispersion of the mass distribution depends on both virial and stellar mass. In order to do this, we consider that the mass distribution is contained within an ellipse whose semi-minor axis is approximately equal to the maximum value of the intrinsic dispersion of the real sample and whose semi-major axis is approximately equal to the mass interval within which the galaxies of our sample are distributed. Again, as was done in the previous case, the distribution of the 100 000 dots is built around the straight line with slope equal to 1 and with intercept on the origin. This means that the global behaviour of the virial mass vs. the stellar mass of the galaxies is similar (no dark matter), although, the intrinsic dispersion depends on the values of both masses. Figure 6 presents this simulated distribution.


\begin{figure}
   \centering
   \includegraphics[angle=0,width=8cm]{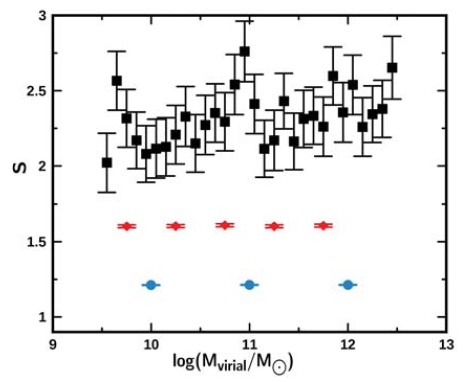}
      \caption{Behaviour of the $BCES_{Bis}$ fit slope for virial mass intervals of the same width from the rectangular simulation. Each point corresponds to the mean value of the mass of the galaxies contained in each mass interval analysed (see Table 3). Circles, diamonds and squares represent intervals of width 1.0, 0.5, and 0.1 in the abscissa.}
         \label{FigVibStab}
   \end{figure}


\begin{figure}
   \centering
   \includegraphics[angle=0,width=8cm]{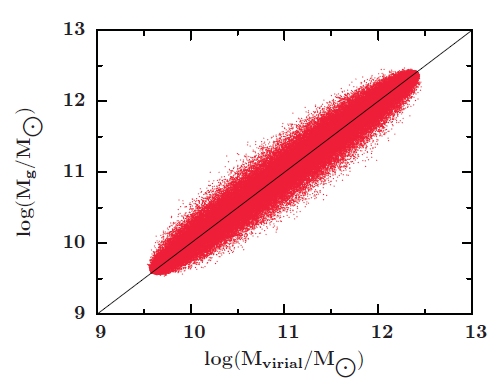}
      \caption{Distribution of the virial vs. stellar mass of the ellipsoidal simulation. The solid line is the one-to-one line.}
         \label{FigVibStab}
   \end{figure}

In Table 4 and Figure 7 we present the coefficients of the linear fits for samples in increasing mass intervals (lower mass cut-off). Given that the total sample was forced to distribute itself around the straight line with unit slope, then the subsamples must also distribute themselves along this same straight line, however, it is possible to see that the value of the slope changes progressively as we consider a larger number of less massive galaxies. The slope only approaches 1 when the mass interval is relatively wide. The slope differs from the expected value by less than 10\% when the interval in the logarithm of the virial mass is approximately equal to 1.5 or less.

In Table 5 we show the results of the analysis that considers mass intervals of the same width at different values of the virial mass. The structure of Table 5 is equal to that of Table 3. In this table we may see how the value of the intrinsic dispersion gets closer to the expected value as the mass interval becomes narrower. If we calculate the intrinsic dispersion at constant mass, its value is  identical to the expected one.


\begin{table*}

\begin{minipage}[t]{\columnwidth}

\renewcommand{\footnoterule}{}  

\caption{Coefficients of the $BCES_{Bis}$ fit for the ellipsoidal simulation of galaxies in increasing-mass-intervals (lower mass cut-off).}

\vskip1.0cm

\begin{tabular}{cccccc}

\hline
$MI$   \footnote{Mass interval within which the galaxies are distributed.}         & $N$  \footnote{Number of galaxies in the mass interval.} & $S$ \footnote{Slope of the $BCES_{Bis}$ fit .}              & $A$   \footnote{Zero point of the $BCES_{Bis}$ fit.}     & $\sigma_{mass}$ \footnote{Intrinsic dispersion of the distribution.}\\

\hline


\multicolumn{5}{c}{}\\
\multicolumn{5}{c}{Intervals that progressively increase their width in amount of 0.1}\\
\multicolumn{5}{c}{}\\

$12.4 \geq M > 12.5$ & 820.00 & -4.0550 $\pm$ 0.3840 & 50.290 $\pm$ 4.7620 & 0.0490 \\
$12.3 \geq M > 12.5$ & $4.2800 \times 10^{3}$  & 1.7950 $\pm$ 0.0360 & -9.8590  $\pm$ 0.4480 & 0.078 \\
$12.2 \geq M > 12.5$ & $7.7590 \times 10^{3}$ & 1.5670 $\pm$ 0.0140 & -7.0130  $\pm$ 0.1780 & 0.108 \\
$12.1 \geq M > 12.5$ & $1.1360 \times 10^{4}$ & 1.4080 $\pm$ $7.6870 \times 10^{-3}$ & -5.0260 $\pm$ 0.0950 & 0.136 \\
$12.0 \geq M > 12.5$ & $1.4900 \times 10^{4}$ & 1.3260 $\pm$ $5.2150 \times 10^{-3}$ & -4.0060  $\pm$ 0.0640 & 0.164 \\
$11.9 \geq M > 12.5$ & $1.8430 \times 10^{4}$ & 1.2640 $\pm$ $3.7760 \times 10^{-3}$ & -3.2310 $\pm$ 0.0460 & 0.192 \\
$11.8 \geq M > 12.5$ & $2.1840 \times 10^{4}$ & 1.2170 $\pm$ $2.9170 \times 10^{-3}$ & -2.6590  $\pm$ 0.0360 & 0.218 \\
$11.7 \geq M > 12.5$ & $2.5340 \times 10^{4}$ & 1.1790 $\pm$ $2.2980 \times 10^{-3}$ & -2.1770  $\pm$ 0.0280 & 0.246 \\
$11.6 \geq M > 12.5$ & $2.8890 \times 10^{4}$ & 1.1530 $\pm$ $1.8700 \times 10^{-3}$ & -1.8610  $\pm$ 0.0230 & 0.274 \\
$11.5 \geq M > 12.5$ & $3.2410 \times 10^{4}$ & 1.1310 $\pm$ $1.5350 \times 10^{-3}$ & -1.5900  $\pm$ 0.0190 & 0.301 \\
$11.4 \geq M > 12.5$ & $3.6000 \times 10^{4}$ & 1.1100 $\pm$ $1.2820 \times 10^{-3}$ & -1.3260  $\pm$ 0.0160 & 0.328 \\
$11.3 \geq M > 12.5$ & $3.9600 \times 10^{4}$ & 1.0970 $\pm$ $1.0840 \times 10^{-3}$ & -1.1670  $\pm$ 0.0130 & 0.357 \\
$11.2 \geq M > 12.5$ & $4.3090 \times 10^{4}$ & 1.0850 $\pm$ $9.3570 \times 10^{-4}$ & -1.0270  $\pm$ 0.0110 & 0.383 \\
$11.1 \geq M > 12.5$ & $4.6600 \times 10^{4}$ & 1.0750 $\pm$ $8.0740 \times 10^{-4}$ & -0.9050  $\pm$ $9.7290 \times 10^{-3}$ & 0.411 \\
$11.0 \geq M > 12.5$ & $5.0170 \times 10^{4}$ & 1.0650 $\pm$ $7.0530 \times 10^{-4}$ & -0.7790  $\pm$ $8.4800 \times 10^{-3}$ & 0.438 \\
$10.9 \geq M > 12.5$ & $5.3670 \times 10^{4}$ & 1.0570 $\pm$ $6.2100 \times 10^{-4}$ & -0.6860  $\pm$ $5.4510 \times 10^{-3}$ & 0.465 \\
$10.8 \geq M > 12.5$ & $5.7250 \times 10^{4}$ & 1.0500 $\pm$ $5.4410 \times 10^{-4}$ & -0.5950  $\pm$ $6.5130 \times 10^{-3}$ & 0.492 \\
$10.7 \geq M > 12.5$ & $6.0740 \times 10^{4}$ & 1.0430 $\pm$ $4.8230 \times 10^{-4}$ & -0.5140  $\pm$ $5.7590 \times 10^{-3}$ & 0.518 \\
$10.6 \geq M > 12.5$ & $6.4150 \times 10^{4}$ & 1.0370 $\pm$ $4.3100 \times 10^{-4}$ & -0.4400  $\pm$ $3.1350 \times 10^{-3}$ & 0.544 \\
$10.5 \geq M > 12.5$ & $6.7700 \times 10^{4}$ & 1.0310 $\pm$ $3.8400 \times 10^{-4}$ & -0.3700  $\pm$ $4.5650 \times 10^{-3}$ & 0.572 \\
$10.4 \geq M > 12.5$ & $7.1210 \times 10^{4}$ & 1.0260 $\pm$ $3.4250 \times 10^{-4}$ & -0.3080 $\pm$ $4.0600 \times 10^{-3}$ & 0.598 \\
$10.3 \geq M > 12.5$ & $7.4690 \times 10^{4}$ & 1.0220 $\pm$ $3.0630 \times 10^{-4}$ & -0.2580 $\pm$ $3.6230 \times 10^{-3}$ & 0.625 \\
$10.2 \geq M > 12.5$ & $7.8160 \times 10^{4}$ & 1.0180 $\pm$ $2.7460 \times 10^{-4}$ & -0.2130 $\pm$ $3.2410 \times 10^{-3}$ & 0.652 \\
$10.1\geq M > 12.5$ & $8.1670 \times 10^{4}$ & 1.0150 $\pm$ $2.4650 \times 10^{-4}$ & -0.1730 $\pm$ $2.9040 \times 10^{-3}$ & 0.679 \\
$10.0\geq M > 12.5$ & $8.5200 \times 10^{4}$ & 1.0110 $\pm$ $2.2140 \times 10^{-4}$ & -0.1320 $\pm$ $2.6020 \times 10^{-3}$ & 0.706 \\
$9.9 \geq M > 12.5$ & $8.8640 \times 10^{4}$ & 1.0080 $\pm$ $1.9960 \times 10^{-4}$ & -0.0960 $\pm$ $2.3400 \times 10^{-3}$ & 0.733 \\
$9.8 \geq M > 12.5$ & $9.2130 \times 10^{4}$ & 1.0050 $\pm$ $1.8010 \times 10^{-4}$ & -0.0610 $\pm$ $2.1060 \times 10^{-3}$ & 0.76 \\
$9.7 \geq M > 12.5$ & $9.5700 \times 10^{4}$ & 1.0030 $\pm$ $1.6230 \times 10^{-4}$ & -0.0310 $\pm$ $1.8930 \times 10^{-3}$ & 0.787 \\
$9.6 \geq M > 12.5$ & $9.9240 \times 10^{4}$ & 1.0000 $\pm$ $1.4670 \times 10^{-4}$ & $-4.4030 \times 10^{-3}$ $\pm$ $1.7060 \times 10^{-3}$ & 0.815 \\
$9.5 \geq M > 12.5$ & $1.0000 \times 10^{5}$ & 1.0000 $\pm$ $1.4360 \times 10^{-4}$ & $2.6620 \times 10^{-3}$ $\pm$ $1.6680 \times 10^{-3}$ & 0.821 \\



\multicolumn{5}{c}{}\\
\multicolumn{5}{c}{Intervals that progressively increase their width in amount of 0.5}\\

\multicolumn{5}{c}{}\\

$12.0 \geq M > 12.5$ & $1.4900 \times 10^{4}$   & 1.3260 $\pm$ $5.2150 \times 10^{-3}$ & -4.0060 $\pm$ 0.0640 & 0.164 \\
$11.5 \geq M > 12.5$ & $3.2410 \times 10^{4}$   & 1.1310 $\pm$ $1.5350 \times 10^{-3}$ & -1.5900 $\pm$ 0.0190 & 0.301 \\
$11.0 \geq M > 12.5$ & $5.0170 \times 10^{4}$   & 1.0650 $\pm$ $7.0530 \times 10^{-4}$ & -0.7790 $\pm$ $8.4800 \times 10^{-3}$ & 0.438 \\
$10.5 \geq M > 12.5$ & $6.7700 \times 10^{4}$   & 1.0310 $\pm$ $3.8400 \times 10^{-4}$ & -0.3700 $\pm$ $4.5650 \times 10^{-3}$ & 0.572 \\
$10.0 \geq M > 12.5$ & $8.5200 \times 10^{4}$   & 1.0110 $\pm$ $2.2140 \times 10^{-4}$ & -0.1320 $\pm$ $2.6020 \times 10^{-3}$ & 0.706 \\
$9.5 \geq M > 12.5$ & $1.0000 \times 10^{5}$ & 1.0000 $\pm$ $1.4360 \times 10^{-4}$ & $2.6620 \times 10^{-3}$ $\pm$ $1.6680 \times 10^{-3}$ & 0.821 \\


\multicolumn{5}{c}{}\\
\multicolumn{5}{c}{Intervals that progressively increase their width in amount of 1.0} \\

\multicolumn{5}{c}{}\\

$11.5 \geq M > 12.5$  & $3.2410 \times 10^{4}$  &  1.1310 $\pm$ $1.5350 \times 10^{-3}$ & -1.5900 $\pm$ 0.0190 & 0.301 \\
$10.5 \geq M > 12.5$  & $6.7700 \times 10^{4}$  &  1.0310 $\pm$ $3.8400 \times 10^{-4}$ & -0.3700 $\pm$ $4.5650 \times 10^{-3}$ & 0.572 \\
$9.5 \geq M > 12.5$  & $1.0000 \times 10^{5}$  &  1.0000 $\pm$ $1.4360 \times 10^{-4}$ & $2.6620 \times 10^{-3}$ $\pm$ $1.6680 \times 10^{-3}$ & 0.821 \\


\hline

\end{tabular}
\end{minipage}
\end{table*}


\begin{table*}

\begin{minipage}[t]{\columnwidth}

\renewcommand{\footnoterule}{}  

\caption{Coefficients of the $BCES_{Bis}$ fit for the ellipsoidal simulation of galaxies in narrow-mass-intervals.}

\vskip1.0cm

\begin{tabular}{cccccc}

\hline
$MI$   \footnote{Mass interval within which the galaxies are distributed.}         & $N$  \footnote{Number of galaxies in the mass interval.} & $S$ \footnote{Slope of the $BCES_{Bis}$ fit .}              & $A$   \footnote{Zero point of the $BCES_{Bis}$ fit.}     & $\sigma_{mass}$ \footnote{Intrinsic dispersion of the distribution.}\\

\hline


\multicolumn{5}{c}{}\\
\multicolumn{5}{c}{Intervals of width 0.1}\\
\multicolumn{5}{c}{}\\

$9.5 \geq M > 9.6$ & 761.00 & -2.1940 $\pm$ 0.2990 & 30.659 $\pm$ 2.8710 & 0.046 \\
$9.6 \geq M > 9.7$ & $3.5420 \times 10^{3}$  & 1.9010 $\pm$ 0.0560 & -8.6660  $\pm$ 0.5370 & 0.046 \\
$9.7 \geq M > 9.8$ & $3.5710 \times 10^{3}$ & 2.3020 $\pm$ 0.0890 & -12.668  $\pm$ 0.8710 & 0.110 \\
$9.8 \geq M > 9.9$ & $3.4870 \times 10^{3}$ & 2.3040 $\pm$ 0.1170 & -12.819 $\pm$ 1.1520 & 0.133 \\
$9.9 \geq M > 10.0$ & $3.4450 \times 10^{3}$ & 2.4300 $\pm$ 0.1350 & -14.205  $\pm$ 1.3470 & 0.148 \\
$10.0 \geq M > 10.1$ & $3.5260 \times 10^{3}$ & 2.4040 $\pm$ 0.1460 & -14.085 $\pm$ 1.4650 & 0.162 \\
$10.1 \geq M > 10.2$ & $3.5080 \times 10^{3}$ & 2.2150 $\pm$ 0.1570 & -12.317 $\pm$ 1.5920 & 0.175 \\
$10.2 \geq M > 10.3$ & $3.4750 \times 10^{3}$ & 2.3440 $\pm$ 0.1720 & -13.759  $\pm$ 1.7610 & 0.182 \\
$10.3 \geq M > 10.4$ & $3.4780 \times 10^{3}$ & 2.1710 $\pm$ 0.1760 & -12.112  $\pm$ 1.8210 & 0.187 \\
$10.4 \geq M > 10.5$ & $3.5070 \times 10^{3}$ & 2.4400 $\pm$ 0.1800 & -15.030  $\pm$ 1.8790 & 0.196 \\
$10.5 \geq M > 10.6$ & $3.5460 \times 10^{3}$ & 2.3180 $\pm$ 0.1900 & -13.888  $\pm$ 2.0020 & 0.204 \\
$10.6 \geq M > 10.7$ & $3.4200 \times 10^{3}$ & 2.2860 $\pm$ 0.1980 & -13.644  $\pm$ 2.1050 & 0.206 \\
$10.7 \geq M > 10.8$ & $3.4860 \times 10^{3}$ & 2.6320 $\pm$ 0.1940 & -17.538  $\pm$ 2.0820 & 0.205 \\
$10.8 \geq M > 10.9$ & $3.5780 \times 10^{3}$ & 2.0400 $\pm$ 0.1910 & -11.284  $\pm$ 2.0740 & 0.209 \\
$10.9 \geq M > 11.0$ & $3.4970 \times 10^{3}$ & 2.4950 $\pm$ 0.2010 & -16.373  $\pm$ 2.2050 & 0.212 \\
$11.0 \geq M > 11.1$ & $3.5690 \times 10^{3}$ & 2.3480 $\pm$ 0.2000 & -14.900  $\pm$ 2.2090 & 0.212 \\
$11.1 \geq M > 11.2$ & $3.5150 \times 10^{3}$ & 2.3630 $\pm$ 0.1960 & -15.210 $\pm$ 2.1880 & 0.209 \\
$11.2 \geq M > 11.3$ & $3.4880 \times 10^{3}$ & 2.4040 $\pm$ 0.1950 & -15.807  $\pm$ 2.1940 & 0.209 \\
$11.3 \geq M > 11.4$ & $3.5980 \times 10^{3}$ & 2.2150 $\pm$ 0.1860 & -13.804  $\pm$ 2.1160 & 0.203 \\
$11.4 \geq M > 11.5$ & $3.5980 \times 10^{3}$ & 2.3530 $\pm$ 0.1870 & -15.498  $\pm$ 2.1390 & 0.202 \\
$11.5 \geq M > 11.6$ & $3.5190 \times 10^{3}$ & 2.1410 $\pm$ 0.1800 & -13.189 $\pm$ 2.0760 & 0.193 \\
$11.6 \geq M > 11.7$ & $3.5460 \times 10^{3}$ & 2.2860 $\pm$ 0.1770 & -15.001 $\pm$ 2.0630 & 0.192 \\
$11.7 \geq M > 11.8$ & $3.5040 \times 10^{3}$ & 2.3710 $\pm$ 0.1660 & -16.119 $\pm$ 1.9480 & 0.181 \\
$11.8 \geq M > 11.9$ & $3.5490 \times 10^{3}$ & 2.3250 $\pm$ 0.1600 & -15.723 $\pm$ 1.8900 & 0.174 \\
$11.9 \geq M > 12.0$ & $3.5230 \times 10^{3}$ & 2.2090 $\pm$ 0.1440 & -14.466 $\pm$ 1.7260 & 0.160 \\
$12.0 \geq M > 12.1$ & $3.5470 \times 10^{3}$ & 2.2320 $\pm$ 0.1290 & -14.872 $\pm$ 2.5570 & 0.146 \\
$12.1 \geq M > 12.2$ & $3.5990 \times 10^{3}$ & 2.3020 $\pm$ 0.1290 & -15.841 $\pm$ 1.3320 & 0.127 \\
$12.2 \geq M > 12.3$ & $3.5790 \times 10^{3}$ & 2.2330 $\pm$ 0.0910 & -15.130 $\pm$ 1.1110 & 0.111 \\
$12.3 \geq M > 12.4$ & $3.4600 \times 10^{3}$ & 2.0700 $\pm$ 0.0560 & -13.249 $\pm$ 0.6880 & 0.0810 \\
$12.4 \geq M > 12.5$ & $3.2990 \times 10^{3}$ & 2.6500 $\pm$ 0.2090 & -20.547 $\pm$ 2.6040 & 0.207 \\



\multicolumn{5}{c}{}\\
\multicolumn{5}{c}{Intervals of width 0.5}\\

\multicolumn{5}{c}{}\\

$9.5 \geq M > 10.0$ & $1.4810 \times 10^{4}$   & 1.3370 $\pm$ $5.3230 \times 10^{-3}$ & -3.2730 $\pm$ 0.0520 & 0.165 \\
$10.0 \geq M > 10.5$ & $1.7490 \times 10^{4}$   & 1.5070 $\pm$ $7.9550 \times 10^{-3}$ & -5.1760 $\pm$ 0.0810 & 0.228 \\
$10.5 \geq M > 11.0$ & $1.7530 \times 10^{4}$   & 1.5940 $\pm$ 0.0100 & -6.3750 $\pm$ 0.1090 & 0.248 \\
$11.0 \geq M > 11.5$ & $1.7770 \times 10^{4}$   & 1.6120 $\pm$ $9.9240 \times 10^{-3}$ & -6.8890 $\pm$ 0.1120 & 0.250 \\
$11.5 \geq M > 12.0$ & $1.7500 \times 10^{4}$   & 1.5090 $\pm$ $7.9400 \times 10^{-3}$ & -5.9950 $\pm$ 0.0930 & 0.228 \\
$12.0 \geq M > 12.5$ & $1.4900 \times 10^{4}$   & 1.3260 $\pm$ $5.2150 \times 10^{-3}$ & -4.0060 $\pm$ 0.0640 & 0.164 \\


\multicolumn{5}{c}{}\\
\multicolumn{5}{c}{Intervals of width 1.0} \\

\multicolumn{5}{c}{}\\

$9.5 \geq M > 10.5$  & $3.2300 \times 10^{4}$  &  1.1310 $\pm$ $1.5410 \times 10^{-3}$ & -1.2900 $\pm$ 0.0150 & 0.302 \\
$10.5 \geq M > 11.5$  & $3.5300 \times 10^{4}$  &  1.2020 $\pm$ $2.2230 \times1 0^{-3}$ & -2.2270 $\pm$ 0.0240 & 0.349 \\
$11.5 \geq M > 12.5$  & $3.2410 \times 10^{4}$  &  1.1310 $\pm$ $1.5350 \times 10^{-3}$ & -1.5900 $\pm$ 0.0190 & 0.301 \\


\hline

\end{tabular}
\end{minipage}
\end{table*}

In Figure 8 we present the behaviour of the slope of the linear fit as a function of the logarithm of the virial mass of the galaxies (same width intervals). The logarithm of the plotted virial mass corresponds to the mean value in each interval of mass. This graph shows the slopes taken from Table 5. We show the slopes for intervals of width equal to 1 (blue dots), the slopes for intervals of width equal to 0.5 (red diamonds) and for intervals with width equal to 0.1 (black squares). Here we may notice that the values of the slopes we get for mass intervals of the same width depend on the virial mass and, in all cases, move away considerably from the expected value ($m=1$). We also notice that when the mass interval is relatively narrow the dependence on virial mass disappears but the value found for it is very different from the expected value.

\end{itemize}


\begin{figure}
   \centering
   \includegraphics[angle=0,width=8cm]{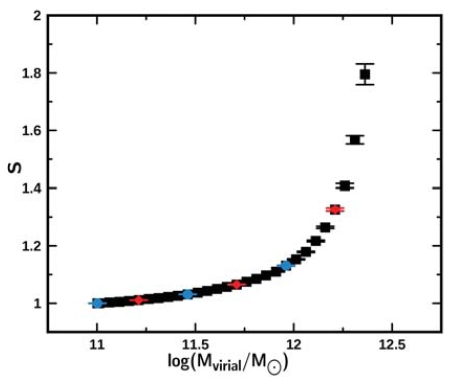}
      \caption{Behaviour of the $BCES_{Bis}$ fit slope in increasing virial mass intervals (lower mass cut-off) from the ellipsoidal simulation. Each point corresponds to the mean value of the mass of the galaxies contained in each mass interval analysed (see Table 4). Circles, diamonds and squares represent intervals that progressively increase their width in amounts of 1.0, 0.5, and 0.1 in the abscissa.}
         \label{FigVibStab}
   \end{figure}


\begin{figure}
   \centering
   \includegraphics[angle=0,width=8cm]{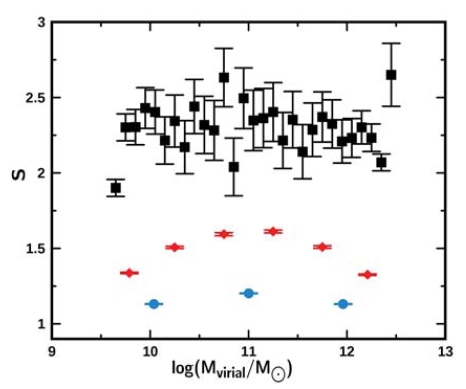}
      \caption{Behaviour of the $BCES_{Bis}$ fit slope for virial mass intervals of the same width from the ellipsoidal simulation. Each point corresponds to the mean value of the mass of the galaxies contained in each mass interval analysed (see Table 5). Circles, diamonds and squares represent intervals of width 1.0, 0.5, and 0.1 in the abscissa.}
         \label{FigVibStab}
   \end{figure}


\begin{figure}
   \centering
   \includegraphics[angle=0,width=8cm]{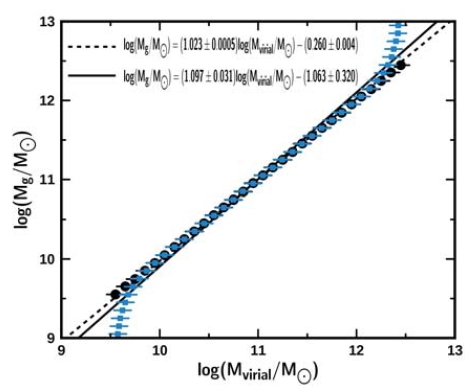}
      \caption{Distribution of the mean values of the virial and stellar mass from the rectangular simulation. The black dots represent the mean values of the stellar mass at quasi-constant virial mass and the blue squares represent the mean values of the virial mass at quasi-constant stellar mass. The continuous line corresponds to the $BQ$ fit and the dotted line corresponds to the $WBQ$ fit.}
         \label{FigVibStab}
   \end{figure}


\begin{figure}
   \centering
   \includegraphics[angle=0,width=8cm]{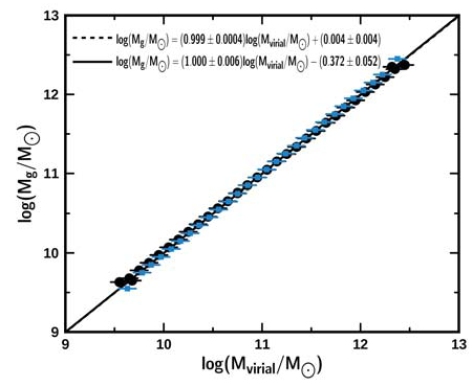}
      \caption{Distribution of the mean values of the virial and stellar mass from the ellipsoidal simulation. The black dots represent the mean values of the stellar mass at quasi-constant virial mass and the blue squares represent the mean values of the virial mass at quasi-constant stellar mass. The continuous line corresponds to the $BQ$ fit and the dotted line corresponds to the $WBQ$ fit.}
         \label{FigVibStab}
   \end{figure}


\begin{figure}
   \centering
   \includegraphics[angle=0,width=8cm]{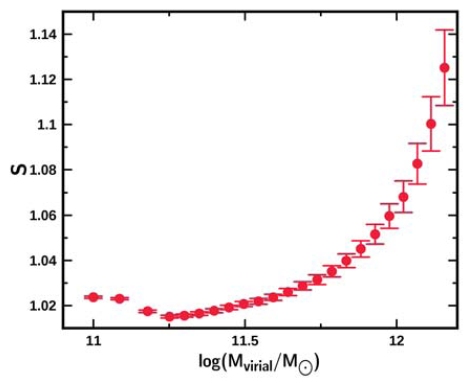}
      \caption{Behaviour of the $WBQ$ fit slope in increasing virial mass intervals (lower mass cut-off) from the distribution of the mean values of masses (rectangular simulation, see Figure 9 and Table 6). Each point corresponds to the mean value of the mass of the galaxies contained in each mass interval analysed.}
         \label{FigVibStab}
   \end{figure}


\begin{figure}
   \centering
   \includegraphics[angle=0,width=8cm]{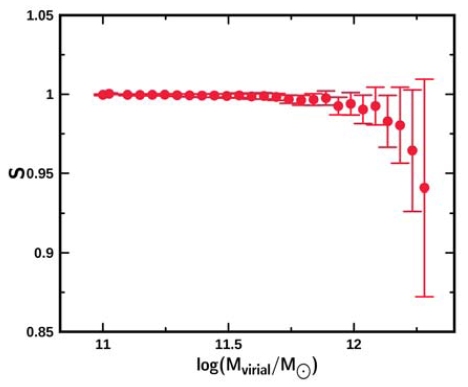}
      \caption{Behaviour of the $WBQ$ fit slope in increasing virial mass intervals (lower mass cut-off) from the distribution of the mean values of the masses (ellipsoidal simulation, see Figure 10 and Table 7). Each point corresponds to the mean value of the mass of the galaxies contained in each mass interval analysed.}
         \label{FigVibStab}
   \end{figure}

Everything we have presented so far confirms that the geometry of the mass distribution of galaxies depends both on the intrinsic properties of the galaxies, as well as on the biases introduced by making arbitrary cuts to the samples. The effect of the different biases introduced is that the values of the slope obtained from the linear fits depend on the mass distribution. We can conclude that the set of calculated linear fit slopes is not an adequate set of parameters to perform analysis of intrinsic properties of the galaxies. This method is, therefore, not useful for investigating the presence of dark matter inside ETGs. In what follows we shall propose another method to do this.

On the other hand, the analysis we have performed confirms that the intrinsic dispersion at quasi-constant magnitude does not depend on the geometric effect, as is the case for other structural relations (see \citealt{nig10}; \citealt{nig11}). This suggests that the intrinsic dispersion at quasi-constant magnitude might be a good tool for analysing the intrinsic properties of galaxies as functions of other variables, such as wavelength, redshift or environment.


\begin{table*}

\begin{minipage}[t]{\columnwidth}

\renewcommand{\footnoterule}{}  

\caption{Coefficients of the $WBQ$ fit in increasing virial mass intervals (lower mass cut-off) from the distribution of the mean values of masses (rectangular simulation, see Figure 9).}

\vskip1.0cm

\begin{tabular}{cccccc}

\hline

$MI$   \footnote{Mass interval within which the galaxies are distributed.}         & $N$  \footnote{Number of points in the mass interval.} & $S$ \footnote{Slope of the $WBQ$ fit.}              & $A$   \footnote{Zero point of the $WBQ$ fit.}\\

\multicolumn{4}{c}{}\\

\multicolumn{4}{c}{Intervals that progressively increase their width in amount of 0.1}\\

\multicolumn{4}{c}{}\\


$12.4  \geq M > 12.5$ & 4.0 & -7.3780  $\pm$  -61.920 & 104.00 $\pm$  872.00 \\
$12.3  \geq M > 12.5$ & 8.0 & 0.4280  $\pm$  0.6820 & 7.1600  $\pm$  9.5500 \\
$12.2  \geq M > 12.5$ & 11 & 1.1510  $\pm$  0.2245 & -1.8000  $\pm$  -2.3600 \\
$12.1  \geq M > 12.5$ & 13 & 1.2410  $\pm$  0.1143 & -2.9100  $\pm$  -1.1500 \\
$12.0  \geq M > 12.5$ & 15 & 1.2210  $\pm$  0.0629 & -2.6600  $\pm$  -0.6340 \\
$11.9  \geq M > 12.5$ & 17 & 1.1810  $\pm$  0.0380 & -2.1800  $\pm$  -0.3840 \\
$11.8  \geq M > 12.5$ & 19 & 1.1510  $\pm$  0.0244 & -1.8000  $\pm$  -0.2460 \\
$11.7  \geq M > 12.5$ & 21 & 1.1250  $\pm$  0.0168 & -1.4800  $\pm$  -0.1690 \\
$11.6  \geq M > 12.5$ & 23 & 1.1000  $\pm$  0.0120 & -1.1800  $\pm$  -0.1220 \\
$11.5  \geq M > 12.5$ & 25 & 1.0820  $\pm$  0.0089 & -0.9700  $\pm$  -0.0915 \\
$11.4  \geq M > 12.5$ & 27 & 1.0680  $\pm$  0.0069 & -0.7950  $\pm$  -0.0702 \\
$11.3  \geq M > 12.5$ & 29 & 1.0590  $\pm$  0.0054 & -0.6920  $\pm$  -0.0551 \\
$11.2  \geq M > 12.5$ & 31 & 1.0510  $\pm$  0.0044 & -0.5950  $\pm$  -0.0443 \\
$11.1  \geq M > 12.5$ & 33 & 1.0450  $\pm$  0.0036 & -0.5170  $\pm$  -0.0363 \\
$11.0  \geq M > 12.5$ & 35 & 1.0390  $\pm$  0.0029 & -0.4540 $\pm$  -0.0301 \\
$10.9  \geq M > 12.5$ & 37 & 1.0350  $\pm$  0.0025 & -0.3970  $\pm$  -0.0253 \\
$10.8  \geq M > 12.5$ & 39 & 1.0310  $\pm$  0.0021 & -0.3540  $\pm$  -0.0214 \\
$10.7  \geq M > 12.5$ & 41 & 1.0280  $\pm$  0.0018 & -0.3210  $\pm$  -0.0183 \\
$10.6  \geq M > 12.5$ & 43 & 1.0260  $\pm$  0.0016 & -0.2890  $\pm$  -0.0158 \\
$10.5  \geq M > 12.5$ & 45 & 1.0230  $\pm$  0.0014 & -0.2610  $\pm$  -0.0137 \\
$10.4  \geq M > 12.5$ & 47 & 1.0210  $\pm$  0.0012 & -0.2400  $\pm$  -0.0120 \\
$10.3  \geq M > 12.5$ & 49 & 1.0200  $\pm$  0.0011 & -0.2260  $\pm$  -0.0106 \\
$10.2  \geq M > 12.5$ & 51 & 1.0190  $\pm$  0.0009 & -0.2080  $\pm$  -0.0094 \\
$10.1  \geq M > 12.5$ & 53 & 1.0170  $\pm$  0.0009 & -0.1910  $\pm$  -0.0084 \\
$10.0  \geq M > 12.5$ & 55 & 1.0160  $\pm$  0.0008 & -0.1770  $\pm$  -0.0075 \\
$9.9  \geq M > 12.5$ & 57 & 1.0150  $\pm$  0.0007 & -0.1660  $\pm$  -0.0067 \\
$9.8  \geq M > 12.5$ & 59 & 1.0150  $\pm$  0.0006 & -0.1600  $\pm$  -0.0060 \\
$9.7  \geq M > 12.5$ & 62 & 1.0170  $\pm$  0.0006 & -0.1880  $\pm$  -0.0053 \\
$9.6  \geq M > 12.5$ & 66 & 1.0230  $\pm$  0.0005 & -0.2520  $\pm$  -0.0048 \\
$9.5  \geq M > 12.5$ & 70 & 1.0230  $\pm$  0.0005 & -0.2600  $\pm$  -0.0046 \\
\end{tabular}
\end{minipage}
\end{table*}


\begin{table*}

\begin{minipage}[t]{\columnwidth}

\renewcommand{\footnoterule}{}  

\caption{Coefficients of the $WBQ$ fit in increasing virial mass intervals (lower mass cut-off) from the distribution of the mean values of masses (ellipsoidal simulation, see Figure 10).}

\vskip1.0cm

\begin{tabular}{cccccc}

\hline

$MI$   \footnote{Mass interval within which the galaxies are distributed.}         & $N$  \footnote{Number of points in the mass interval.} & $S$ \footnote{Slope of the $WBQ$ fit .}              & $A$   \footnote{Zero point of the $WBQ$ fit.}\\

\multicolumn{4}{c}{}\\

\multicolumn{4}{c}{Intervals that progressively increase their width in amount of 0.1}\\

\multicolumn{4}{c}{}\\


$12.4  \geq M > 12.5$ & 0.0 & 0.0000   $\pm$  0.0000 & 0.0000 $\pm$  0.0000  \\
$12.3  \geq M > 12.5$ & 4.0 & 0.3253   $\pm$  0.4095 & 8.3300 $\pm$  6.6800  \\
$12.2  \geq M > 12.5$ & 6.0 & 0.8429   $\pm$  0.1471 & 1.9300 $\pm$  1.7000   \\
$12.1  \geq M > 12.5$ & 8.0 & 0.9409   $\pm$  0.0687 & 0.7230  $\pm$ 0.7510  \\
$12.0  \geq M >  12.5$ & 10 & 0.96440    $\pm$ 0.0383 & 0.4330  $\pm$ 0.4120  \\
$11.9  \geq M >  12.5$ & 12 & 0.9804    $\pm$ 0.0241 & 0.2370  $\pm$ 0.2550  \\
$11.8  \geq M >  12.5$ & 14 & 0.9829    $\pm$ 0.0164 & 0.2070  $\pm$ 0.1730  \\
$11.7  \geq M >  12.5$ & 16 & 0.9925    $\pm$ 0.0119 & 0.0894  $\pm$ 0.1250  \\
$11.6  \geq M >  12.5$ & 18 & 0.9904    $\pm$ 0.0090 & 0.1150  $\pm$ 0.0939  \\
$11.5  \geq M >  12.5$ & 20 & 0.9939    $\pm$ 0.0070 & 0.0723  $\pm$ 0.0730  \\
$11.4  \geq M >  12.5$ & 22 & 0.9925    $\pm$ 0.0056 & 0.0893  $\pm$ 0.0574  \\
$11.3  \geq M >  12.5$ & 24 & 0.9976    $\pm$ 0.0045 & 0.0285  $\pm$ 0.0465  \\
$11.2  \geq M >  12.5$ & 26 & 0.9966    $\pm$ 0.0037 & 0.0396  $\pm$ 0.0378  \\
$11.1  \geq M >  12.5$ & 28 & 0.9962    $\pm$ 0.0031 & 0.0446  $\pm$ 0.0315  \\
$11.0  \geq M >  12.5$ & 30 & 0.9969    $\pm$ 0.0026 & 0.0370  $\pm$ 0.0265  \\
$10.9  \geq M >  12.5$ & 32 & 0.9983    $\pm$ 0.0022 & 0.0200  $\pm$ 0.0225  \\
$10.8  \geq M >  12.5$ & 34 & 0.9989    $\pm$ 0.0019 & 0.0128  $\pm$ 0.0193  \\
$10.7  \geq M >  12.5$ & 36 & 0.9986    $\pm$ 0.0017 & 0.0163  $\pm$ 0.0166  \\
$10.6  \geq M >  12.5$ & 38 & 0.9993    $\pm$ 0.0015 & 0.0087  $\pm$ 0.0144  \\
$10.5  \geq M >  12.5$ & 40 & 0.9989    $\pm$ 0.0013 & 0.0127  $\pm$ 0.0125  \\
$10.4  \geq M >  12.5$ & 42 & 0.9992    $\pm$ 0.0011 & 0.0093  $\pm$ 0.0110  \\
$10.3  \geq M >  12.5$ & 44 & 0.9991    $\pm$ 0.0009 & 0.0099  $\pm$ 0.0097  \\
$10.2  \geq M >  12.5$ & 46 & 0.9993    $\pm$ 0.0009 & 0.0081  $\pm$ 0.0086  \\
$10.1  \geq M >  12.5$ & 48 & 0.9994    $\pm$ 0.0008 & 0.0071  $\pm$ 0.0076  \\
$10.0  \geq M >  12.5$ & 50 & 0.9997    $\pm$ 0.0007 & 0.0032  $\pm$ 0.0068  \\
$9.9  \geq M >  12.5$& 52 &   0.9996    $\pm$ 0.0006 & 0.0041  $\pm$ 0.0060  \\
$9.8  \geq M >  12.5$ & 54 &  0.9996   $\pm$ 0.0006 & 0.0052   $\pm$0.0054  \\
$9.7  \geq M >  12.5$ & 56 &  0.9996    $\pm$ 0.0005 & 0.0042  $\pm$ 0.0048  \\
$9.6  \geq M >  12.5$ & 59 &  1.0003    $\pm$ 0.0004 & -0.0035  $\pm$-0.0043  \\
$9.5  \geq M >  12.5$ & 60 &  0.9996    $\pm$ 0.0004 & 0.0044  $\pm$ 0.0042  \\
\end{tabular}
\end{minipage}
\end{table*}

\subsection{Investigating the presence of dark matter inside ETGs. Case 2}

In the previous section we have seen that the intrinsic dispersion of the mass distribution of galaxies at quasi-constant mass does not depend on the
geometric effect. We propose that we could use the value of this intrinsic dispersion as a tool to investigate the possible presence of dark matter inside ETGs. But, in which way could we achieve this ?

We  note first, as seen in section 7.2, that the mass distribution has two interesting properties, one is its width and the other is the mean value of the distribution. In what follows we shall analyse the relationship of these properties with dark matter.

a) The width of the distribution (intrinsic dispersion) as a function of mass (virial or stellar). This property helps us to study the physical origin of the intrinsic dispersion, given that this width depends on the virial and stellar mass of galaxies (see Figure 1), it could also depend on other variables such as wavelength, redshift and/or environment. If the behaviour of this dispersion as a function of virial mass is different from its behaviour as a function of the stellar mass, then the difference would be related to the presence of dark matter. In a forthcoming paper we shall study this relation in depth.

b) The mean value of the distribution as a function of mass (virial or stellar). This property may help us carry out a first approach to the study of the dark matter inside ETGs given that we may perform a linear fit to the mean values of the stellar mass as a function of the virial mass and viceversa. Implementing this method we must consider that the dispersion depends on the virial and stellar mass (see Figure 1).
In order to account for this fact, it is necessary to calculate the mean value of the distribution of both variables and perform a linear regression. The linear regression obtained in this way may be used
as an alternative method to investigate the global properties of the ETGs sample, in particular to investigate whether or not there is dark matter
inside ETGs. At this point we must emphasize that the application of this method constitutes a first approach to the study of the dark matter inside ETGs. This study must be supplemented by a fuller study in which other variables are taken into consideration; variables such as wavelength, redshift and/or the environment that might affect the determination of the amount of dark matter present.

In what follows when we express the idea of quasi-constant mass, we shall mean mass intervals in the logarithm of width equal to 0.1.

We propose two different linear regressions to analyse the distribution of masses: the first one is a bisector fit applied to the mean value of the distribution at quasi-constant mass ($BQ$ fit), and the second one is a weighted bisector fit applied to the mean value of the distribution at quasi-constant mass ($WBQ$ fit). In the last case, the weight used to perform the fit was the number of galaxies inside the mass interval.

To investigate the degree of confidence which we can give the methods we are proposing, we have used the simulations presented in the previous section. The results may be seen in Figures 9 and 10 for the rectangular and ellipsoidal simulations respectively. In Figures 9 and 10 we plot the mean values of the distribution of the stellar mass at quasi-constant virial mass (black dots), the mean values of the distribution of the virial mass at quasi-constant stellar mass (blue squares) and two different fits, the first one corresponds to the $BQ$ fit (continuous line) and the second one corresponds to the $WBQ$ fit (dotted line). From the Figures 9 and 10, we can see that the $WBQ$ fit produces the best estimation of the parameters of the simulated samples.

But, is the $WBQ$ fit affected by the geometric effect? In Tables 6 and 7 and Figures 11 and 12 we present the results of the values of the slope obtained from the $WBQ$ fit to subsamples of dots (lower mass cut-off) where it is clearly shown that the geometric effect is practically nonexistent. The slope approaches 1 even when the mass interval is relatively narrow. The slope differs from the expected value by more than 10\% when the interval in the logarithm of the virial mass is less than 0.3 for the ellipsoidal simulation and 0.8 for the rectangular simulation. We also can see that the error in determining the slope is relatively small when we have a wide interval of masses. This error, in the worst case, is of the order of 2\% so this will be the error that we consider due to the fit method.

It has been proven that our alternative method, the $WBQ$ fit, allows the recovery, in a trustworthy fashion, of the global behaviour of the masses of galaxies from the simulated samples. In  section 5 we  apply this method to the analysis of the galaxies in our real DR9 samples.

\label{lastpage}


\begin{thebibliography}{}


\bibitem[\protect\citeauthoryear{Abazajian et al.} {2009}]{aba09} Abazajian et al. 2009, ApJS, 182, 543

\bibitem[\protect\citeauthoryear{Aihara et al.} {2011}]{aih11} Aihara, H., Allende Prieto, C., An, D. et al. 2011, ApJS, 193, 29

\bibitem[\protect\citeauthoryear{Akritas \& Bershady} {1996}]{akr96} Akritas, M. G., \& Bershady, M. A. 1996, ApJ, 470, 706

\bibitem[\protect\citeauthoryear{Auger et al.} {2010}]{aug10} Auger, M. W., Treu, T., Gavazzi, R., Bolton, A. S., Koopmans, L. V. E., \& Marshall, P. J. 2010, ApJ, 721, 163	 

\bibitem[\protect\citeauthoryear{Auger et al.} {2010a}]{aug10a} Auger, M. W., Treu, T., Bolton, A. S., Gavazzi, R., Koopmans, L. V. E., Marshall, P. J., Moustakas, L. A., \& Burles, S. 2010a, ApJ, 724, 511

\bibitem[\protect\citeauthoryear{Barnab\`e et al.} {2011}]{bar11} Barnab\`e, M, Czoske, O, Koopmans, L. V. E., Treu, T., \& Bolton, A. S. 2011, MNRAS, 415, 2215

\bibitem[\protect\citeauthoryear{Bell et al.} {2003}]{bel03} Bell, E. F., McIntosh, D. H.,  Katz N., \& Weinberg M. D. 2003, ApJS, 149, 289


\bibitem[\protect\citeauthoryear{Bernardi et al.} {2003a}]{ber03a} Bernardi, M., et al. 2003a, AJ, 125, 1817

\bibitem[\protect\citeauthoryear{Bernardi et al.} {2003b}]{ber03b} Bernardi, M., et al. 2003b, AJ, 125, 1849

\bibitem[\protect\citeauthoryear{Bertin \& Stiavelli} {1993}]{ber93} Bertin,G . \& Stiavelli, M. 1993, Rep. Prog. Phys., 56, 493



\bibitem[\protect\citeauthoryear{Brinchmann et al.} {2004}]{bri04} Brinchmann, J., S. Charlot, S., White,S. D. M., Tremonti, C., Kauffmann, G.. Heckman, T. \& Brinkmann, J. 2004, MNRAS, 351, 1151

\bibitem[\protect\citeauthoryear{Burbidge \& Burbidge} {1975}]{bur75} Burbidge, E. M. \& Burbidge, G. R., 1975, in Stars and Stellar Systems, vol IX, Galaxies and the Universe, 81


\bibitem[\protect\citeauthoryear{Cappellari et al.} {2006}]{cap06} Cappellari, M., Bacon, R., Bureau, R., Damen, M.C., Davies, R.L., de Zeeuw, P.T., Emsellem, E., Falc\'on-Barroso, J., Krajnovi\'c, D., Kuntscher, H., McDermid, R.M., Peletier, R.F., Sarzi, M., van den Bosch, R.C.E. \& van de Ven, G. 2006, MNRAS, 366, 1126

\bibitem[\protect\citeauthoryear{Cappellari et al} {2006}] {cappellari2006} Cappellari, M., Bacon, R., Bureau, M., et al., 2006, MNRAS, 366, 1126

\bibitem[\protect\citeauthoryear{Cappellari et al.} {2012}]{cap12} Cappellari, M., McDermid, R. M. Alatalo, K et al., G. 2012, Nature, 484, 485

\bibitem[\protect\citeauthoryear{Cappellari et al.} {2013}]{cap13} Cappellari, M., Scott, N., Alatalo, K., et al. 2013, MNRAS, 432, 1709

\bibitem[\protect\citeauthoryear{Chabrier} {2003}]{cha03}
Chabrier, G. 2003, PASP, 115, 763

\bibitem[\protect\citeauthoryear{Chen et al.} {2010}]{che10} Chen, C. W., Cot\'e, P., West, A. A., Peng, E. W. \& Ferrarese, L.  2010, ApJS, 191, 1

\bibitem[\protect\citeauthoryear{Das et al.} {2011}]{das11} Das, P., Gerhard, O., M\'endez, R. H., Teodorescu, A. M., \& de Lorenzi, F. 2011, MNRAS, 415, 1244

\bibitem[\protect\citeauthoryear{Dutton et al.} {2013}]{dut13} Dutton, A. A., Macci\'o, A. V., Mendel, J. T., Simard, L., \& de Lorenzi, F. 2013, MNRAS, 432, 2496

\bibitem[\protect\citeauthoryear{De Bruyne et al.} {2001}]{deb01} De Bruyne V., Dejonghe, H., Pizzela, A., Bernardi, M., \& Zeilinger, W. W. 2001, ApJ, 546, 903

\bibitem[\protect\citeauthoryear{Emsellem et al.} {2004}]{ems04} Emsellem, E., Cappellari, M., Peletier, R. F., McDermid, R. M., Bacon, R., Bureau, M., Copin, Y., Davies, R. L., Krajnovic, Davor., Kuntschner, H., et al. 2004, MNRAS, 352, 721



\bibitem[\protect\citeauthoryear{Gavazzi et al.} {2007}]{gav07}  Gavazzi, R., Treu, T., Rhodes, J. D., Koopmans, L. V. E., Bolton, A. S., Burles, S., Massey, R. J., \& Moustakas, L. A. 2007, ApJ, 667, 176

\bibitem[\protect\citeauthoryear{Gebhardt et al.} {2001}] {gebhardt2001} Gebhardt, K., Lauer, T. R., Kormendy, J., et al. 2001, AJ, 122, 2469

\bibitem[\protect\citeauthoryear{Gerhard et al.} {2001}] {gerhard2001} Gerhard, O., Kronewitter, A., Saglia, R. P., and Bender, R., 2001, AJ, 121, 1936


\bibitem[\protect\citeauthoryear{Graham et al.} {2005}]{gra05} 	Graham, A. W., Driver, S. P., Petrosian, V., Conselice, C. J., Bershady, M. A., Crawford, S. M.,  and Goto, T. 2005, AJ, 130, 1535

\bibitem[\protect\citeauthoryear{Graves \& Faber} {2010}]{gra10} Graves, G. J., \& Faber, S. M. 2010, ApJ, 717, 803

\bibitem[\protect\citeauthoryear{Hern\'andez et al.} {2010}]{her10} Hern\'andez, X.; Mendoza, S.; Suarez, T.; Bernal, T. 2010, Invisible Universe: Proceedings of the Conference. AIP Conference Proceedings, 1241, 918

\bibitem[\protect\citeauthoryear{Hern\'andez et al.} {2012}]{her12} Hern\'andez, X., Jim\'enez, M. A., \& Allen, C. 2012, European Physical Journal C, 72, 1884

\bibitem[\protect\citeauthoryear{Hoekstra et al.} {2005}]{hoe05} Hoekstra, H., Hsieh, B. C., Yee, H. K. C., Lin, H., Gladders, M. D. 2005, ApJ, 635, 73


\bibitem[\protect\citeauthoryear{Humphrey et al.} {2006}]{hum06} Humphrey, P. J., Buote, D. A., Gastaldello, F., Zappacosta, L., Bullock, J. S. et al. 2006, ApJ, 646, 899

\bibitem[\protect\citeauthoryear{Hyde \& Bernardi} {2009}]{hyd09} Hyde, J. B., \& Bernardi, M. 2009, MNRAS, 394, 1978

\bibitem[\protect\citeauthoryear{Isobe et al.} {1990}]{iso90} Isobe, T., Feigelson, E. D., Akritas, M. G., \& Babu, G. J.  1990, ApJ, 364, 104

\bibitem[\protect\citeauthoryear{J\o rgensen et al.} {1995a}]{jor95a} J\o rgensen, I.,  Franx, M., \& K{\ae}rgaard, p. 1995, MNRAS, 273, 1097

\bibitem[\protect\citeauthoryear{J\o rgensen et al.} {1995b}]{jor95b} J\o rgensen, I.,  Franx, M., \& K{\ae}rgaard, p. 1995, MNRAS, 276, 1341


\bibitem[\protect\citeauthoryear{Jiang \& Kochanek} {2007}]{jia07} Jiang, G., \& Kochanek, C. S. 2007, ApJ, 671, 1568

\bibitem[\protect\citeauthoryear{Jim\'enez et al.} {2013}]{jim13} Jim\'enez, N.A., Garc\'{\i}a, G., Hern\'andez, X. \& Nasser, L. 2013, ApJ, 768, 142

\bibitem[\protect\citeauthoryear{Kauffmann et al.} {2003}]{kau03} Kauffmann, G., Heckman, T. M., White, S. D. M. 2003, MNRAS, 341, 33

\bibitem[\protect\citeauthoryear{Koopmans et al.} {2006}]{koopmans2006} Koopmans, L.V.E., Treu, T., Bolton, A.S., et al, 2006, ApJ, 649, 599

\bibitem[\protect\citeauthoryear{Koopmans \& Treu} {2010}]{koo10} Koopmans, L.V.E. \& Treu, T. 2010, HiA, 15, 61

\bibitem[\protect\citeauthoryear{Kroupa} {2001}]{kro01} Kroupa, P. 2001, MNRAS, 322, 231

\bibitem[\protect\citeauthoryear{Kroupa et al.} {2010}]{kro10} Kroupa, P., Famaey, B., de Boer, K. S., Dabringhausen, J., Pawlowsky, M. S., Boily, C. M., Jerjen, H., Forbes, D., Hensler, G., \& Metz, M. 2010, A\&A, 523, A32


\bibitem[\protect\citeauthoryear{Lagattuta et al.} {2010}]{lag10} Lagattuta, D. J., Fassnacht, C. D., Auger, M. W., Marshall, P. J., Bradac, M., Treu, T., Gavazzi, R., Schrabback, T., Faure, C., \& Anguita, T. 2010, ApJ, 716, 1579

\bibitem[\protect\citeauthoryear{Lintott et al.} {2008}]{lin08} Lintott, C. J., Schawinski, K., Slosar, A., Land, K.,
Bamford, S,. Thomas, D., Raddick, M. J., Nichol, R. C., Szalay, A., Andreescu, D., Murray, P., \& van den Berg J. 2008, MNRAS, 389, 1179

\bibitem[\protect\citeauthoryear{Magorrian et al} {1998}] {magorrian1998} Magorrian, J., Tremaine, S., Richstone, D., et al, 1998, AJ, 115,2285

\bibitem[\protect\citeauthoryear{McGaugh \& Wolf} {2010}]{mcg10} McGaugh, S. S., \& Wolf, J. 2010, ApJ, 722, 248

\bibitem[\protect\citeauthoryear{Milgrom} {1983}]{mil83} Milgrom, M. 1983, ApJ, 270, 365



\bibitem[\protect\citeauthoryear{Napolitano et al.} {2010}]{nap10} Napolitano, N. R., Romanowsky, A. J., \& Tortora, C. 2010, MNRAS, 405, 2351

\bibitem[\protect\citeauthoryear{Navarro et al.} {1996}]{nav96} Navarro, J. F., Frenk, C. S., \& White, S. D. M. 1996, ApJ, 462, 563



\bibitem[\protect\citeauthoryear{Nigoche-Netro et al.} {2008}]{nig08} Nigoche-Netro, A., Ruelas-Mayorga, A., \& Franco-Balderas, A. 2008, A\&A, 491, 731

\bibitem[\protect\citeauthoryear{Nigoche-Netro et al.} {2009}]{nig09} Nigoche-Netro, A., Ruelas-Mayorga, A., \& Franco-Balderas, A. 2009, MNRAS, 392, 1060

\bibitem[\protect\citeauthoryear{Nigoche-Netro et al.} {2010}]{nig10} Nigoche-Netro, A., Aguerri, J. A. L., Lagos, P., Ruelas-Mayorga, A., S\'anchez, L. J., \& Machado, A. 2010, A\&A, 516, 96

\bibitem[\protect\citeauthoryear{Nigoche-Netro et al.} {2011}]{nig11} Nigoche-Netro, A., Aguerri, J. A. L., Lagos, P., Ruelas-Mayorga, A., S\'anchez, L. J., Mu\~noz-Tu\~n\'on, C., \& Machado, A. 2011, A\&A, 534, 61

\bibitem[\protect\citeauthoryear{Robertson et al.} {2006}]{rob06}Robertson, B., Cox, T. J., Hernquist, L., Franx, M., Hopkins, P. F., Martini,P., \& Springel, V. 2006, ApJ, 641, 21

\bibitem[\protect\citeauthoryear{Romanowsky et al.} {2003}]{rom03} Romanowsky, A. J., Douglas, N. G., Arnaboldi, M., Kuijken, K., Merrifield M. R., Napolitano,N. R., Capaccioli, M., \& Freeman, K. C. 2003, Science, 301, 1696




\bibitem[\protect\citeauthoryear{Sanders \& McGaugh} {2002}]{san02} Sanders, R. H.,\& McGaugh, S. S. 2002, ARA\&A, 40, 263

\bibitem[\protect\citeauthoryear{Schulz et al.} {2010}]{sch10} Schulz, A. E., Mandelbaum, R., \& Padmanabhan, Nikhil. 2010, MNRAS, 408, 1463

\bibitem[\protect\citeauthoryear{Simon \& Geha} {2007}]{sim07} Simon, J. D. \& Geha, M. 2007, ApJ, 670, 313

\bibitem[\protect\citeauthoryear{Sofue \& Rubin} {2001}]{sof01} Sofue, Y \& Rubin, V. 2001, ARA\&A, 39, 137

\bibitem[\protect\citeauthoryear{Sonnenfeld et al.} {2012}]{son12} Sonnenfeld, A., Treu, T., Gavazzi, R., Marshall, P. J., Auger, M. W., Suyu, S. H., Koopmans, L. V. E., \& Bolton, A. S. 2012, ApJ, 752, 163

\bibitem[\protect\citeauthoryear{Spinrad \& Peimbert} {1975}]{spi75} Spinrad, H \& Peimbert M. 1975, in Stars and Stellar Systems, vol. IX, Galaxies and the Universe, 37

\bibitem[\protect\citeauthoryear{Teodorescu et al.} {2011}]{teo11} Teodorescu, A. M., M\'endez, R. H., Bernardi, F., Thomas, J., Das, P., \& Gerhard, O. 2011, ApJ, 736, 65

\bibitem[\protect\citeauthoryear{Thomas et al.} {2007}]{thomas2007} Thomas, J., Saglia, R. P., Bender, R., et al, 2007, MNRAS, 382, 657

\bibitem[\protect\citeauthoryear{Thomas et al.} {2011}]{thomas2011} Thomas, J., Saglia, R. P., Bender, R., et al, 2011, MNRAS, 415, 545


\bibitem[\protect\citeauthoryear{Tortora et al.} {2009}]{tor09} Tortora, C., Napolitano, N. R., Romanowsky, A. J., Capaccioli, M., \& Covone, G. 2009, MNRAS, 396, 1132

\bibitem[\protect\citeauthoryear{Treu \& Koopmans} {2004}]{tre04} Treu, T., \& Koopmans, L. V. E. 2004, ApJ, 611, 739

\bibitem[\protect\citeauthoryear{Treu} {2010}]{tre10} Treu, T. 2010, ARA\&A, 48, 87


\bibitem[\protect\citeauthoryear{Tremonti et al.} {2004}] {trm04} Tremonti, C. A., Heckman, T. M., Kauffmann, G., Brinchmann, J., Charlot, S., White, S. D. M., Seibert, M., Peng, E. W., Schlegel, D. J., Uomoto, A., Fukugita, M., \& Brinkmann, J. 2004, ApJ, 613, 898

\bibitem[\protect\citeauthoryear{van der Marel} {1991}] {vandermarel1991} van der Marel, R. P., 1991, MNRAS, 253, 710

\bibitem[\protect\citeauthoryear{Williams et al.} {2009}] {williams2009} Williams, M. J., Bureau, M, and Cappellari, M., 2009, MNRAS, 400, 1665

\bibitem[\protect\citeauthoryear{York et al.} {2000}]{yor00} York, D. G. et al. 2000, AJ, 120, 1579


\end{thebibliography}
\end{document}